\documentclass[prl,twocolumn,calc,epsfig]{revtex4-1}

\usepackage{graphics}
\usepackage{amsfonts}
\usepackage{amsmath}
\usepackage{epsfig}
\newif\ifdraft \drafttrue
\catcode`\ä = \active \catcode`\ö = \active \catcode`\ü = \active
\catcode`\Ä = \active \catcode`\Ö = \active \catcode`\Ü = \active
\catcode`\ß = \active \catcode`\é = \active \catcode`\è = \active
\catcode`\ë = \active \catcode`\ô = \active \catcode`\ê = \active
\catcode`\ø = \active \catcode`\ò = \active \catcode`\í = \active
\catcode`\Ó = \active \catcode`\ú = \active \catcode`\á = \active
\catcode`\ã = \active \catcode`\à = \active
\defä{\"a} \defö{\"o} \defü{\"u} \defÄ{\"A} \defÖ{\"O} \defÜ{\"U} \defß{\ss} \defé{\'{e}}
\defè{\`{e}} \defë{\"{e}} \defô{\^{o}} \defê{\^{e}} \defø{\o} \defò{\`{o}} \defí{\'{i}}
\defÓ{\'{O}} \defú{\'{u}} \defá{\'{a}} \defã{\~{a}}\defà{\`{a}}

\begin{document}

\title{From Planar Solitons to Vortex Rings and Lines:\\Cascade of Solitonic Excitations in a Superfluid Fermi Gas}
\author{Mark J.H. Ku,  Biswaroop Mukherjee, Tarik Yefsah, Martin W. Zwierlein}

\affiliation{MIT-Harvard Center for Ultracold Atoms, Research Laboratory of Electronics, and Department of Physics, Massachusetts Institute of Technology, Cambridge, Massachusetts 02139, USA}

\begin{abstract}
We follow the time evolution of a superfluid Fermi gas of resonantly interacting $^6$Li atoms after a phase imprint. Via tomographic imaging, we observe the formation of a planar dark soliton, its subsequent snaking, and its decay into a vortex ring, which in turn breaks to finally leave behind a single solitonic vortex. In intermediate stages we find evidence for an exotic structure resembling the $\Phi$-soliton, a combination of a vortex ring and a vortex line. Direct imaging of the nodal surface reveals its undulation dynamics and its decay via the puncture of the initial soliton plane.  The observed evolution of the nodal surface represents dynamics beyond superfluid hydrodynamics, calling for a microscopic description of unitary fermionic superfluids out of equilibrium.
\end{abstract}

\maketitle

Solitonic excitations such as solitons, vortices and vortex rings are found in a large variety of nonlinear media, from classical fluids and plasmas to polyacetylene chains and superconductors. While ubiquitous, their intrinsic properties are tailored by the host medium. In superfluids, which are characterized by a complex order parameter with a well-defined phase and a non-viscous flow, such excitations correspond to phase defects and exhibit properties non-existent in their classical counterparts. There, a vortex is topologically protected owing to the quantized circulation of the velocity field, and a traveling soliton experiences superfluid back flow determined by the phase difference across it~\cite{Scott2011SolitonDynamics,Efimkin2014BCSsoliton}. The quantum statistics of the particles forming the superfluid is yet another ingredient which dramatically affects the properties of these defects. In Fermi superfluids, as opposed to the bosonic case, dark solitons and vortices are known to host in-gap fermionic excitations in their cores, from the Andreev bound states in the generic case~\cite{caro64bound,Antezza2007FermiSoliton}, to the more exotic Majorana fermions in the presence of spin-orbit coupling~\cite{Xu2014SolitonMajorana,Liu2015SolitonMajorana}.

Importantly, in a quantum fluid with short-ranged interactions, these phase defects are localized within the microscopic length scale of the system: the healing length $\xi$. The healing length sets the length scale above which the superfluid dynamics is well captured by the hydrodynamic formalism. At length scales on the order of $\xi$ or smaller, a microscopic description is required, and this is where the dichotomy between Bose superfluids and Fermi superfluids becomes stringent. While weakly interacting Bose-Einstein condensates (BEC) are well understood in terms of the Gross-Pitaevskii (GP) theory, a complete microscopic wave equation for strongly-interacting Fermi superfluids remains to be established. At the mean-field level, a unified description can be formulated within the Bogoliubov--de Gennes (BdG) formalism, which connects to the GP equation in the limit of weakly interacting BECs, and contains the necessary fermionic degrees of freedom in the Bardeen-Cooper-Schrieffer limit (BCS)~\cite{Antezza2007FermiSoliton,Wen2009,Scott2011SolitonDynamics,Efimkin2014BCSsoliton}. However, while the BdG framework provides a good description of these two limiting cases, it is unclear whether it contains the right ingredients to quantitatively handle the behavior of solitonic excitations in the strongly correlated regime, where the dynamics near the core of these phase defects is highly nontrivial~\cite{bulgac2013TSLDA}. The unitary Fermi gas realized in ultracold atom experiments offers a unique opportunity to clarify this issue, as it resides at the point of the BEC-BCS crossover where beyond mean-field correlations are expected to be the strongest~\cite{Zwerger2011BECBCS}. It is also the regime where the healing length $\xi$ is the smallest -- on the order of the inter-particle spacing -- such that phase defects are as localized as possible in a quantum fluid.

%The experimental challenge in this context is not only to create these excitations in a controlled way, but also to set the conditions for the defects' core dynamics to be revealed. A natural approach for the observation of such microscopic dynamics is the relaxation of a highly energetic defect~\cite{Bulgac2014Vortexrings,Reichl2013VortexRing,scherpelz2014vortexring, wlaz14lifecycle}.
A natural approach to experimentally reveal the core dynamics of such defects is to trigger their decay. Solitonic excitations indeed follow a well-defined hierarchy in terms of stability and energy cost in three dimensions, the planar soliton being the most energetic and unstable towards the formation of other solitary waves~\cite{Muryshev1999Soliton,Feder2000Soliton, brand01svort, Brand2002solitonicvortex, Komineas2003soliton,mateo2014chladni}. In weakly interacting BECs, dark solitons have been observed to decay into vortex rings and vortices~\cite{ande01ring,dutton2001shock,Becker2013SolitonicVortex} as a consequence of the snake instability, the undulation of the soliton plane~\cite{Muryshev1999Soliton}. In the case of strongly interacting Fermi superfluids, similar scenarios have been predicted numerically within a mean-field approximation~\cite{wlaz14lifecycle, Reichl2013VortexRing, scherpelz2014vortexring}, but an experimental support of such microscopic dynamics is still lacking.

%of these excitations can be observed through the interaction of these phase-defects, for instance in vortex recombination where two initially separated vortices come close to each other and combine their cores~\cite{bulgac11dynamics}.

%Complex microscopic dynamics is also expected to take place in the decay of a highly energetic defect.

\begin{figure*}
    \begin{center}
    \includegraphics[width=180mm]{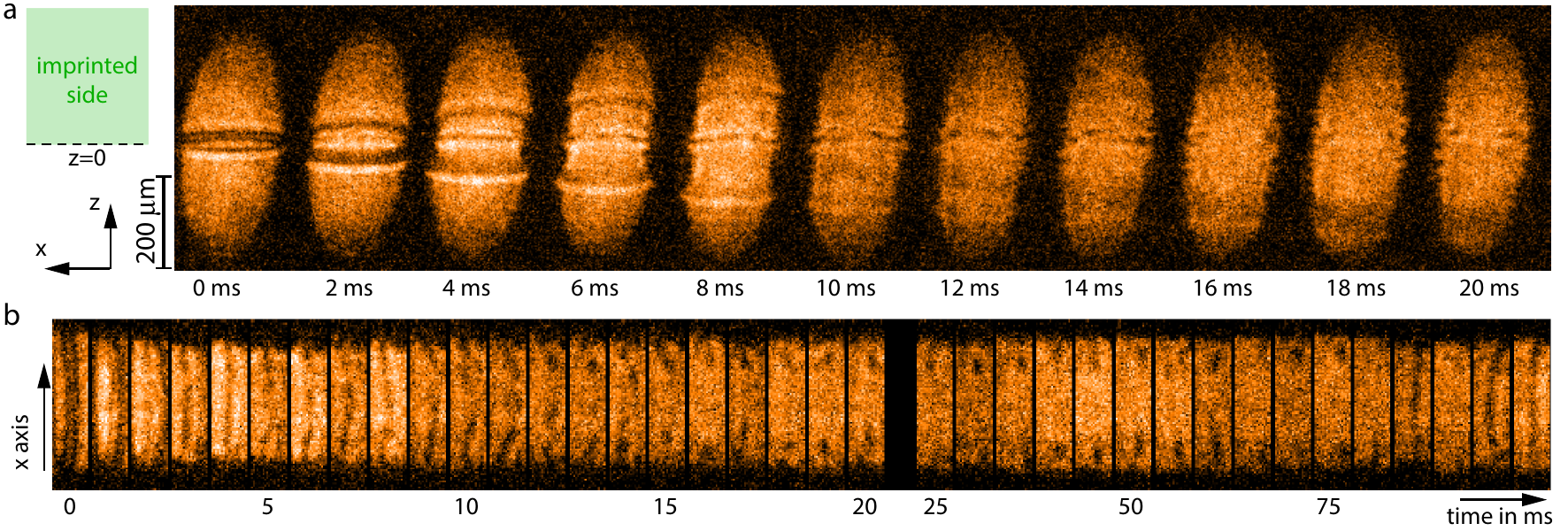}
  \caption{(Color online) (a) Cascade of solitonic excitations in a unitary Fermi superfluid following the phase imprint. A planar soliton snakes and decays into a vortex ring. Shown are images of the density distribution in the central slice of the superfluid, after rapid ramp and time of flight, for the first $20\,\mathrm{ms}$ after the imprint. The imprint also generates two sound waves propagating towards the edges. (b) Time-series of the central slice up to $t=100\,$ms, cropped to the region around $z=0$.}
    \label{f:fig1}
    \end{center}
\end{figure*}

In this Letter, we create a cascade of solitonic excitations in a unitary Fermi gas of $^6$Li atoms. Starting from a planar dark soliton created via phase imprinting, we observe the formation of ring defects which eventually decay into a single solitonic vortex.  By means of a tomographic imaging technique~\cite{Ku2014svortex}, we are able to follow the surface dynamics of the soliton's nodal plane at the level of the interparticle spacing, as it snakes, breaks and converts into the topologically protected solitonic vortex. Our measurements allow for a quantitative analysis of the snaking dynamics of the initial dark soliton, awaiting comparison to time-dependent theories of strongly correlated fermions.

We create a strongly interacting fermionic superfluid using a balanced mixture of the two lowest hyperfine states of $^6$Li ($\left|1\right\rangle$ and $\left|2\right\rangle$) at a Feshbach resonance~\cite{kett08rivista}. Our atomic clouds contain $\sim 7\times 10^5$ atoms per spin state confined in an elongated trap, combining a tight radial optical potential (in the $x$-$y$ plane) and a shallower axial magnetic potential (along the $z$ axis). The axial and radial trapping frequencies are $\omega_z/2\pi = 10.87(1)\,\rm Hz$ and $\omega_\perp/2\pi = 69(6)\,\rm Hz$, respectively. The axial and radial Thomas-Fermi radii of the cloud are $R_z=326(2)\,\mu$m and $R_\perp=54(2)\,\mu$m, and correspond to a chemical potential at the center of the cloud of $\mu=h\times3.7(1)\,\rm kHz = 54(4)\, \hbar \omega_\perp$. The gas is thus deep in the three-dimensional regime. Gravity slightly weakens the trapping potential along the vertical $y$-direction (see~\cite{seeSuppmat}).
Phase imprinting is realized as in Refs.~\cite{burg99soliton,dens00,Becker2008solitons,Yefsah2013Soliton,Ku2014svortex}, whereby one half of the superfluid is exposed to a blue-detuned laser beam for a time sufficient to advance the phase by approximately $\pi$. These experimental parameters are similar to those of previous works~\cite{Yefsah2013Soliton,Ku2014svortex}, where a single solitonic vortex was detected and observed to undergo a deterministic precessional motion for several seconds. Here, we study the evolution of the excitations at early times following the phase imprint. To probe such dynamics, we employ the detection scheme introduced in~\cite{Ku2014svortex}, which combines the so-called rapid ramp technique and tomographic imaging~\cite{seeSuppmat}.
The rapid ramp converts the spatial variations of the order parameter $\Delta(\mathbf{r})$ of the fermion pair superfluid into resolvable density variations of a Bose gas of molecules~\cite{Yefsah2013Soliton,kett08varenna,zwie05vort}. It is a powerful tool for the study of phase defects in atomic Fermi gases, as the presence of any phase singularity manifests itself as a strong density depletion. In our experimental sequence, the rapid ramp is performed at a variable wait time $t$ following the imprint. We then slice a thin layer of the atomic cloud at a chosen $y$ position, and destructively probe its density distribution via absorption imaging.

Fig.~\ref{f:fig1}a shows a time sequence of images recorded in the first 20\,ms after the phase imprint, which correspond to the density distribution at the central slice (near $y=0$) of the superfluid. At the location of the imprinted phase jump ($z=0$), a slow and straight dark soliton emerges and subsequently undergoes a snaking motion, seeding the puncture of the nodal surface. The broken soliton evolves into a vortex ring structure, visible in the central slice as a pair of nodal points. Fig.~\ref{f:fig1}b presents a zoomed-in view of the soliton's time evolution up to 100\,ms after the imprint.
Simultaneous with the soliton's core dynamics, two wavefronts quickly propagate to the edges of the cloud, which we identify as sound waves.
The upper and lower sound wavefronts are found to propagate at speeds of 13.1(4)\,mm/s and 13.1(8)\,mm/s respectively, which coincide with the speed of sound of 12.9(1)\,mm/s estimated from the peak density using the relation $c_s=\sqrt{\xi_{\mathrm{B}}/3}v_{\mathrm F}$, where $\xi_{\mathrm{B}}=0.37$ is the Bertsch parameter~\cite{ku2012thermodynamics} and $v_{\mathrm F}$ the Fermi velocity~\footnote{The observations made here are reminiscent of the first phase imprinting experiment realized in Bose-Einstein condensates~\cite{burg99soliton}, where two waves rapidly propagating in opposite directions were created, one of which has been interpreted as a fast dark soliton. Here, we observe that both wavefronts propagate at the speed of sound.}. The apparent large amplitude of these sound waves is a consequence of the rapid-ramp~~\cite{seeSuppmat}.

\begin{figure}
    \begin{center}
    \includegraphics[width=86mm]{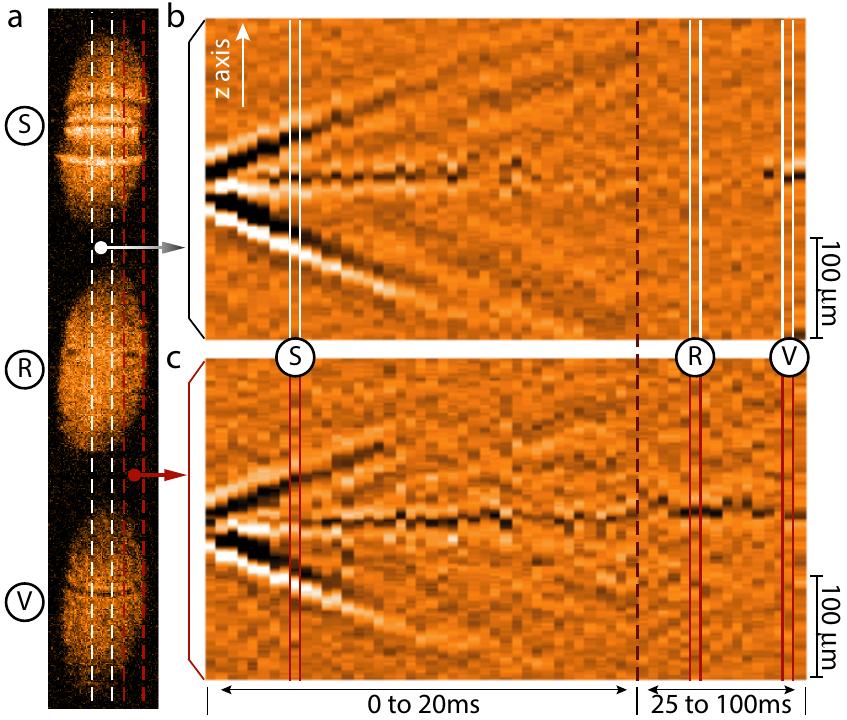}
  \caption{(Color online) Overview of the dynamics following the phase imprinting. (a) Representative images of the central slice at $t=4$, 50 and 95\,ms showing a planar soliton (S), the nodal points of a vortex ring (R) and a vortex line (V). The vertical lines indicate the regions of interest of the generation of residuals in (b). (b) Residuals on the central slice along the central axial cut $x=0$ (upper panel) and an outer axial cut (lower panel). Dark (bright) color indicates density depletion (excess). Two sound waves rapidly propagate to the edges, while a sharp depletion remains at the center. Around $t=5\,$ms, a second set of shallower sound waves is emitted. The residuals show the puncture of the soliton plane in the central slice ($t\sim15\,$ms) and the return of a vortex line at $t\sim80\,$ms.}
    \label{f:fig2}
    \end{center}
\end{figure}

The dynamics is analyzed in detail in Fig.~\ref{f:fig2}, showing residuals of the central slice as a function of time, along the axial cut at $x=0$ (Fig.~\ref{f:fig2}b) and along its outer edge near $x=R_\perp$ (Fig.~\ref{f:fig2}c).
The different characteristic speeds of the various waves generated after the phase imprint are apparent. One recognizes the two initially created sound waves following linear trajectories with opposite slopes, while the dark soliton remains near $z=0$ with negligible velocity. A second set of shallower sound waves is emitted about $\sim5\,$ms after the initial sound wavefronts, forming all together a pattern of hydrodynamic wakes. The rapid vanishing of the sound contrasts with the persistence of the solitonic wave near $z=0$. The unbroken soliton appears in the residual near $x=0$ (Fig.~\ref{f:fig2}b) as well as the outer (Fig.~\ref{f:fig2}c) residuals, while the vortex ring structure is signaled by a localized point-like depletion visible exclusively in the outer residuals. Close to 100 ms after the imprint, a single solitonic vortex remains, precessing in the superfluid, with a period of $\sim1.4\,$s along the $z$ axis~\cite{Yefsah2013Soliton,Ku2014svortex} (see also~\cite{seeSuppmat}), which is more than an order of magnitude longer than the duration of the cascade.

\begin{figure}
    \begin{center}
    \includegraphics[width=86mm]{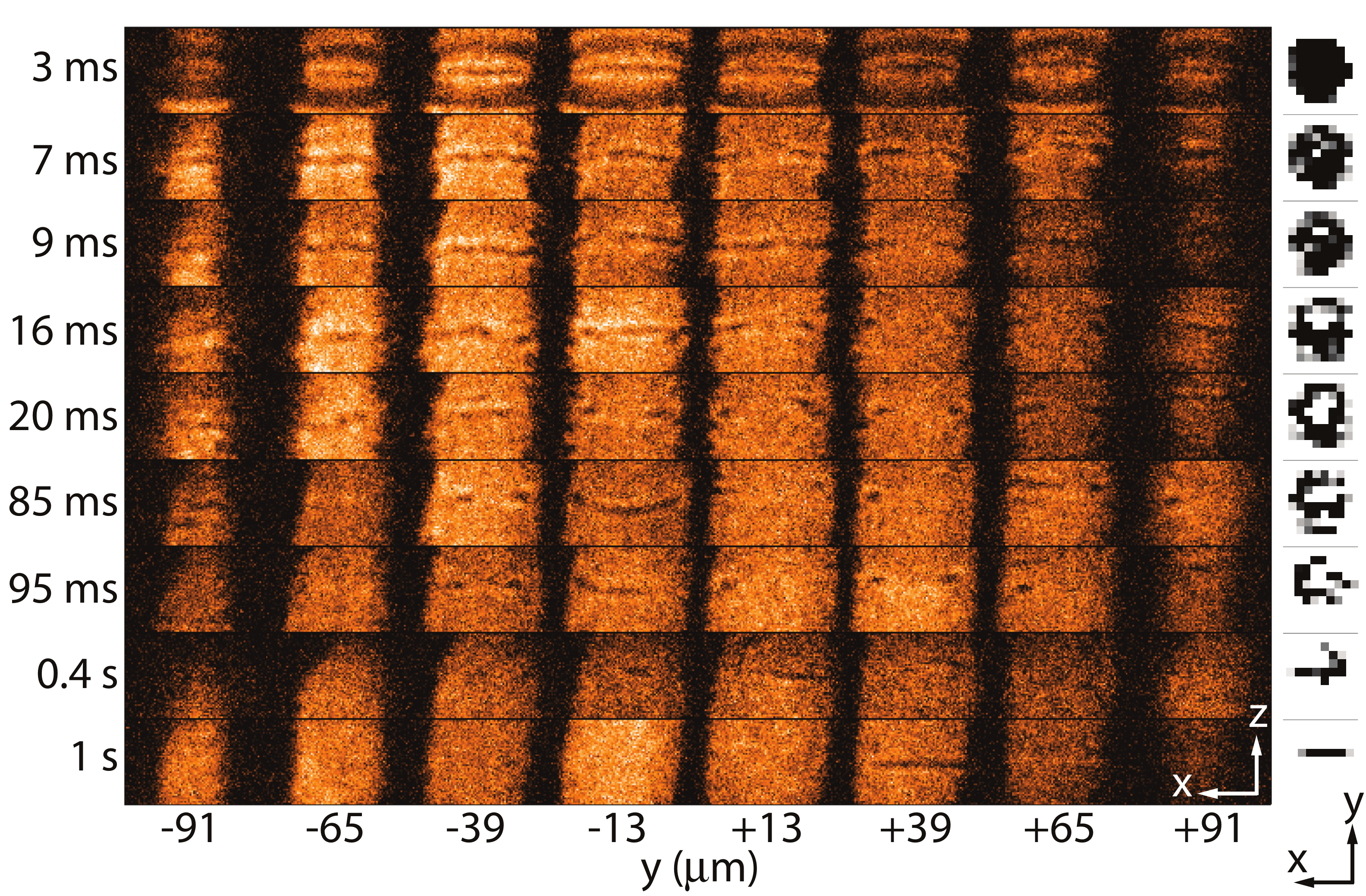}
  \caption{(Color online) Tomography of the cascade. Main panel (left): representative tomographic images at different stages of the cascade. Right column: structure of the depletion due to the defect, as would be seen along the $z$-axis, reconstructed from the tomographic images. $t=3\,$ms: sharp density depletion across the whole cloud signaling a planar dark soliton. $t=7\,$ms: snaking of the soliton plane and first signature of a puncture. $t=9$, 16 and 20\,ms: the puncture in the upper half of the soliton plane grows and yields an asymmetric vortex ring, with a nodal area left at bottom. $t=85\,$ms: the lower nodal area is punctured and a vortex line forms across a vortex ring. $t=95\,$ms and $t=400\,$ms: the ring part of the defect progressively disintegrates. $t=1000\,$ms: a single solitonic vortex remains.}
    \label{f:fig3}
    \end{center}
\end{figure}

In order to obtain a complete picture of the dynamics, three-dimensional tomography of the superfluid is performed after the rapid ramp. Fig.~\ref{f:fig3} displays a set of representative tomographic images at various times, giving access to the local pair density after the ramp. From these images, we are able to reconstruct the structure of the defect engraved in the superfluid and follow its time evolution. The right panel of Fig.~\ref{f:fig3} shows the reconstructed depletion as would be seen from the long axis of the cloud. At early times, a surface of depletion cutting through the entire cloud's section is observed, the planar dark soliton. It subsequently tears in its upper half and then undergoes a cascade into structures with smaller and smaller nodal area. The hole appearing in the initial nodal surface is seen to continuously grow in size, leading to the formation of a transient asymmetric vortex ring, combining the bottom part of the initial soliton and a vortex line bent in a semicircle on the upper part.
At this stage one might anticipate that the nodal area will naturally heal into a standard vortex ring, by shrinking into a single loop with a core of size $\xi$. However, the tomographic images obtained at later times suggest a more complicated scenario~\cite{seeSuppmat}, where a second puncture occurs in the lower nodal plane. This results in the formation of a horizontal line depletion which we interpret as a vortex line intersecting the vortex ring. This structure is seen in the tomographic images e.g. at $t=85\,\mathrm{ms}$ (see Fig.~\ref{f:fig3}) and resembles the $\Phi$-soliton recently proposed in Ref.~\cite{mateo2014chladni}. At even later times, the ring part of this exotic defect progressively disintegrates (as seen at $t=95\,\mathrm{ms}$ and  $400\,\mathrm{ms}$ in Fig.~\ref{f:fig3}), leaving behind a single solitonic vortex ($t=1\,\mathrm{s}$), which precesses in the superfluid. It is the precession of this remnant solitonic vortex which has been studied in~\cite{Yefsah2013Soliton,Ku2014svortex}.

Recently, several theoretical works have treated the evolution of fermionic superfluids following a phase imprint~\cite{sacha14imprint,scherpelz2014vortexring,Scherpelz2014defects,wlaz14lifecycle} and the possible cascade scenarios following the decay of a planar dark soliton~\cite{Bulgac2014Vortexrings,Wen2013DarkSoliton,Reichl2013VortexRing,scherpelz2014vortexring} via various mean-field approaches. In some of these works, it has been found numerically that in a cylindrically symmetric potential with negligible dissipation, a planar soliton decays into a vortex ring, which then undergoes a long-lived oscillatory motion along the $z$ axis~\cite{Bulgac2014Vortexrings,Wen2013DarkSoliton,Reichl2013VortexRing}. By mimicking experimental imperfections, such as trap distortion due to gravity~\cite{wlaz14lifecycle}, and imperfect phase imprinting~\cite{scherpelz2014vortexring,Scherpelz2014defects}, later works found that the vortex ring further decays into a single remnant vortex. The proposed scenarios are however distinct from our observations. Recent simulations based on the Gross-Pitaevskii equation reveal a multitude of dynamical pathways towards the final single vortex, via various intermediate ``Chladni solitons''~\cite{mateo2014chladni,Brandprivatecommunication}.

\begin{figure}
    \begin{center}
    \includegraphics[width=86mm]{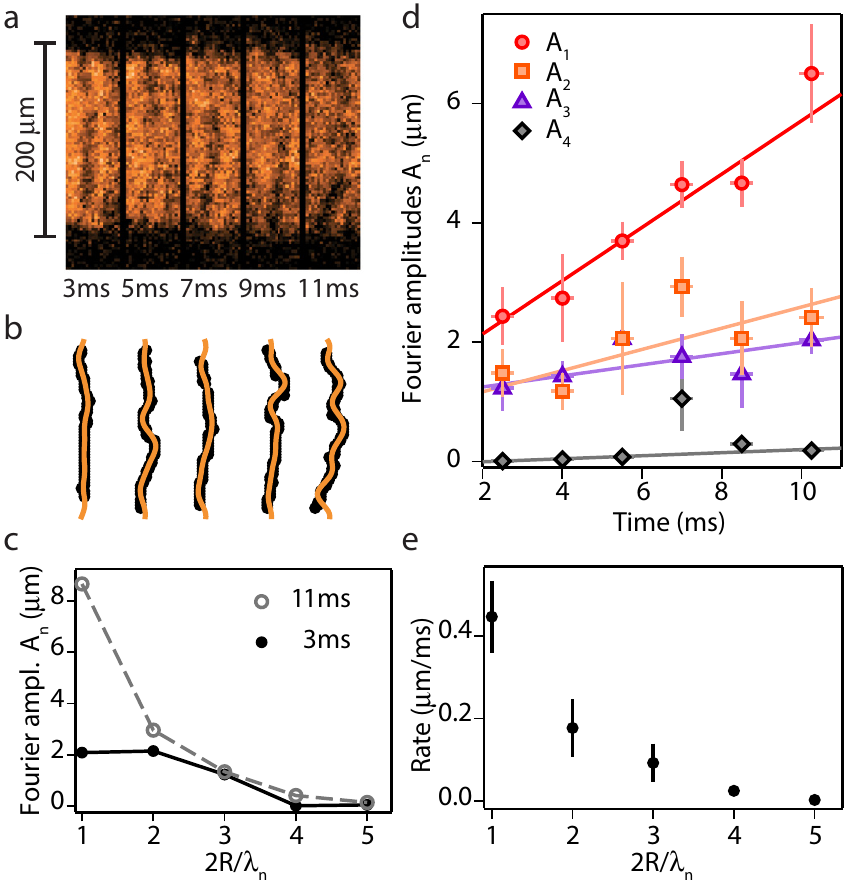}
  \caption{(Color online) Spectral analysis of the snaking dynamics. (a) Snapshots of the soliton's undulation. (b) Extracted undulation shapes $z_{\rm s}(x)$ (black dots) along with the corresponding Fourier expansions to the 5$^{\rm th}$ order (solid gold line). (c) Fourier spectra at $t=3$ and 11\,ms. (d) Fourier amplitudes $A_n$ as a function of time, for $n=1$ (circle), 2 (square), 3 (triangle) and 4 (diamond) and their fits to a line (solid lines). The error bars indicate the standard deviation of the mean obtained from a 3-points binning of the data. (e) Growth rates $\dot A_n$ for $n=1$ to 5, obtained from the linear fits in (d), with the error bars being the fit error.}
    \label{f:fig4}
    \end{center}
\end{figure}

At the origin of the cascade lies the snaking instability of the soliton. In order to quantify this undulation dynamics, we perform a Fourier analysis on the shape of the soliton $z_{\rm s}(x) \approx A_0 + \Sigma_{n=1}^N A_n \cos\left(2\pi n \frac{x}{2R_\perp} + \phi_n\right)$ in terms of the transverse modes of wavelengths $\lambda_n=2R_\perp/n$, the integer $n$ being the mode number, with Fourier amplitudes $A_n$ and phases $\phi_n$. Fig.~\ref{f:fig4}a displays selected images of the snaking soliton, and Fig.~\ref{f:fig4}b the corresponding nodal profiles $z_{\rm s}(x)$ obtained from the density minima. For each of these curves, the result of the Fourier expansion up to the $5^{\rm th}$ order (gold solid line) is superimposed, illustrating that the undulation observed here is well characterized in terms of transverse mode excitations. Fourier spectra of the soliton's undulation are obtained for 2\,ms$\,\leq\,t\,\leq 11$\,ms (see Fig.~\ref{f:fig4}c), and the evolution of the amplitudes $A_n$ is reported in Fig.~\ref{f:fig4}d. We find that the fundamental mode $\lambda_1=2R_\perp$ largely dominates this dynamics, with a relative weight consistently higher than that of the harmonics and a significantly larger growth rate $\dot A_1$. The rates $\dot A_n$ decrease as the mode number $n$ increases (see Fig.~\ref{f:fig4}e). The contribution of the modes $n>5$ was found to be insignificant.

The trend of the growth rates $\dot A_n$ (Fig.~\ref{f:fig4}d,e) is in strong contrast to what is expected for a uniform transverse confinement. In that case, a hydrodynamic argument~\cite{Kamchatnov2008} implies that the growth rate at long wavelength increases linearly with the transverse mode~\footnote{Note that the growth rate that we report here is in the units of a speed, and not a frequency as considered in~\cite{Kamchatnov2008,scott2012solitondecay}.}, as confirmed, for small wavevectors, in BdG numerical simulations for crossover superfluids~\cite{scott2012solitondecay}. Physically, the predominance of the fundamental mode found here is not surprising given the transverse inhomogeneity of the trapped superfluid. Indeed, the phase difference $\Delta\phi$ across the soliton is tied to its relative velocity $v_{\rm s}/c$, where $c$ is the superfluid critical velocity, through a unique current-phase relation $\Delta\phi=f(v_{\rm s}/c)$. While the functional form of $f$ is still under debate for the unitary Fermi gas, it is expected to be monotonic, at least at small relative velocities~\cite{spuntarelli2011soliton,scott2012solitondecay}. Therefore, for an inhomogeneous transverse confinement where $c$ is highest at the center of the cloud ($r=0$) and gradually decreases towards the edges ($r\rightarrow R_\perp$), the phase jump $\Delta\phi$ across a moving soliton can only be maintained if its velocity is locally adjusted such that $v_{\rm s}(r)/c(r)$ is constant. The soliton thus develops a drum-like profile that continuously stretches along the propagation direction, which explains at least in part the growth of $A_1$.

In conclusion, we have observed a cascade of solitonic excitations in a strongly-interacting Fermi superfluid, from an initial planar dark soliton towards a final, remnant solitonic vortex, through an intermediate ring structure resembling the recently predicted $\Phi$-soliton~\cite{mateo2014chladni}. At the origin of the cascade lies the snaking instability, which we characterized quantitatively by studying the evolution of transverse Fourier modes. The breaking dynamics of the unitary Fermi gas studied here occurs at the scale of the interparticle spacing, and provides a novel experimental input for microscopic theories of strongly interacting fermions. A natural extension of our work is to approach a regime where the snake instability is inhibited, e.g. via a strong confining potential in the radial direction. Future prospects are a measurement of the soliton's current-phase relation in the BEC-BCS crossover~\cite{scott2012solitondecay}, the detection and manipulation of Andreev bound states trapped inside the soliton~\cite{caro64bound,Antezza2007FermiSoliton}, and the creation of soliton trains in the presence of spin imbalance, which would realize one limit of Fulde-Ferrell-Larkin-Ovchinnikov states~\cite{Yoshida2007LO, Radz2011LO, Lutchyn2011Soliton}.

We would like to thank Joachim Brand and Lev Pitaevskii for fruitful discussions, Julian Struck for a critical reading of the manuscript, and Parth Patel and Zhenjie Yan for assistance in the data analysis. This work was supported by the NSF, the ARO MURI on Atomtronics, AFOSR PECASE and MURI on Exotic Phases, and the David and Lucile Packard Foundation.

\bibliography{References}

\begin{thebibliography}{44}%
\makeatletter
\providecommand \@ifxundefined [1]{%
 \@ifx{#1\undefined}
}%
\providecommand \@ifnum [1]{%
 \ifnum #1\expandafter \@firstoftwo
 \else \expandafter \@secondoftwo
 \fi
}%
\providecommand \@ifx [1]{%
 \ifx #1\expandafter \@firstoftwo
 \else \expandafter \@secondoftwo
 \fi
}%
\providecommand \natexlab [1]{#1}%
\providecommand \enquote  [1]{``#1''}%
\providecommand \bibnamefont  [1]{#1}%
\providecommand \bibfnamefont [1]{#1}%
\providecommand \citenamefont [1]{#1}%
\providecommand \href@noop [0]{\@secondoftwo}%
\providecommand \href [0]{\begingroup \@sanitize@url \@href}%
\providecommand \@href[1]{\@@startlink{#1}\@@href}%
\providecommand \@@href[1]{\endgroup#1\@@endlink}%
\providecommand \@sanitize@url [0]{\catcode `\\12\catcode `\$12\catcode
  `\&12\catcode `\#12\catcode `\^12\catcode `\_12\catcode `\%12\relax}%
\providecommand \@@startlink[1]{}%
\providecommand \@@endlink[0]{}%
\providecommand \url  [0]{\begingroup\@sanitize@url \@url }%
\providecommand \@url [1]{\endgroup\@href {#1}{\urlprefix }}%
\providecommand \urlprefix  [0]{URL }%
\providecommand \Eprint [0]{\href }%
\providecommand \doibase [0]{http://dx.doi.org/}%
\providecommand \selectlanguage [0]{\@gobble}%
\providecommand \bibinfo  [0]{\@secondoftwo}%
\providecommand \bibfield  [0]{\@secondoftwo}%
\providecommand \translation [1]{[#1]}%
\providecommand \BibitemOpen [0]{}%
\providecommand \bibitemStop [0]{}%
\providecommand \bibitemNoStop [0]{.\EOS\space}%
\providecommand \EOS [0]{\spacefactor3000\relax}%
\providecommand \BibitemShut  [1]{\csname bibitem#1\endcsname}%
\let\auto@bib@innerbib\@empty
%</preamble>
\bibitem [{\citenamefont {Scott}\ \emph {et~al.}(2011)\citenamefont {Scott},
  \citenamefont {Dalfovo}, \citenamefont {Pitaevskii},\ and\ \citenamefont
  {Stringari}}]{Scott2011SolitonDynamics}%
  \BibitemOpen
  \bibfield  {author} {\bibinfo {author} {\bibfnamefont {R.~G.}\ \bibnamefont
  {Scott}}, \bibinfo {author} {\bibfnamefont {F.}~\bibnamefont {Dalfovo}},
  \bibinfo {author} {\bibfnamefont {L.~P.}\ \bibnamefont {Pitaevskii}}, \ and\
  \bibinfo {author} {\bibfnamefont {S.}~\bibnamefont {Stringari}},\ }\href@noop
  {} {\bibfield  {journal} {\bibinfo  {journal} {Phys. Rev. Lett.}\ }\textbf
  {\bibinfo {volume} {106}},\ \bibinfo {pages} {185301} (\bibinfo {year}
  {2011})}\BibitemShut {NoStop}%
\bibitem [{\citenamefont {Efimkin}\ and\ \citenamefont
  {Galitski}(2015)}]{Efimkin2014BCSsoliton}%
  \BibitemOpen
  \bibfield  {author} {\bibinfo {author} {\bibfnamefont {D.~K.}\ \bibnamefont
  {Efimkin}}\ and\ \bibinfo {author} {\bibfnamefont {V.}~\bibnamefont
  {Galitski}},\ }\href {\doibase 10.1103/PhysRevA.91.023616} {\bibfield
  {journal} {\bibinfo  {journal} {Phys. Rev. A}\ }\textbf {\bibinfo {volume}
  {91}},\ \bibinfo {pages} {023616} (\bibinfo {year} {2015})}\BibitemShut
  {NoStop}%
\bibitem [{\citenamefont {Caroli}\ \emph {et~al.}(1964)\citenamefont {Caroli},
  \citenamefont {Gennes},\ and\ \citenamefont {Matricon}}]{caro64bound}%
  \BibitemOpen
  \bibfield  {author} {\bibinfo {author} {\bibfnamefont {C.}~\bibnamefont
  {Caroli}}, \bibinfo {author} {\bibfnamefont {P.~d.}\ \bibnamefont {Gennes}},
  \ and\ \bibinfo {author} {\bibfnamefont {J.}~\bibnamefont {Matricon}},\
  }\href@noop {} {\bibfield  {journal} {\bibinfo  {journal} {Phys. Lett.}\
  }\textbf {\bibinfo {volume} {9}},\ \bibinfo {pages} {307} (\bibinfo {year}
  {1964})}\BibitemShut {NoStop}%
\bibitem [{\citenamefont {Antezza}\ \emph {et~al.}(2007)\citenamefont
  {Antezza}, \citenamefont {Dalfovo}, \citenamefont {Pitaevskii},\ and\
  \citenamefont {Stringari}}]{Antezza2007FermiSoliton}%
  \BibitemOpen
  \bibfield  {author} {\bibinfo {author} {\bibfnamefont {M.}~\bibnamefont
  {Antezza}}, \bibinfo {author} {\bibfnamefont {F.}~\bibnamefont {Dalfovo}},
  \bibinfo {author} {\bibfnamefont {L.~P.}\ \bibnamefont {Pitaevskii}}, \ and\
  \bibinfo {author} {\bibfnamefont {S.}~\bibnamefont {Stringari}},\ }\href@noop
  {} {\bibfield  {journal} {\bibinfo  {journal} {Phys. Rev. A}\ }\textbf
  {\bibinfo {volume} {76}},\ \bibinfo {pages} {043610} (\bibinfo {year}
  {2007})}\BibitemShut {NoStop}%
\bibitem [{\citenamefont {Xu}\ \emph {et~al.}(2014)\citenamefont {Xu},
  \citenamefont {Mao}, \citenamefont {Wu},\ and\ \citenamefont
  {Zhang}}]{Xu2014SolitonMajorana}%
  \BibitemOpen
  \bibfield  {author} {\bibinfo {author} {\bibfnamefont {Y.}~\bibnamefont
  {Xu}}, \bibinfo {author} {\bibfnamefont {L.}~\bibnamefont {Mao}}, \bibinfo
  {author} {\bibfnamefont {B.}~\bibnamefont {Wu}}, \ and\ \bibinfo {author}
  {\bibfnamefont {C.}~\bibnamefont {Zhang}},\ }\href {\doibase
  10.1103/PhysRevLett.113.130404} {\bibfield  {journal} {\bibinfo  {journal}
  {Phys. Rev. Lett.}\ }\textbf {\bibinfo {volume} {113}},\ \bibinfo {pages}
  {130404} (\bibinfo {year} {2014})}\BibitemShut {NoStop}%
\bibitem [{\citenamefont {Liu}(2015)}]{Liu2015SolitonMajorana}%
  \BibitemOpen
  \bibfield  {author} {\bibinfo {author} {\bibfnamefont {X.-J.}\ \bibnamefont
  {Liu}},\ }\href {\doibase 10.1103/PhysRevA.91.023610} {\bibfield  {journal}
  {\bibinfo  {journal} {Phys. Rev. A}\ }\textbf {\bibinfo {volume} {91}},\
  \bibinfo {pages} {023610} (\bibinfo {year} {2015})}\BibitemShut {NoStop}%
\bibitem [{\citenamefont {Wen}\ and\ \citenamefont {Huang}(2009)}]{Wen2009}%
  \BibitemOpen
  \bibfield  {author} {\bibinfo {author} {\bibfnamefont {W.}~\bibnamefont
  {Wen}}\ and\ \bibinfo {author} {\bibfnamefont {G.}~\bibnamefont {Huang}},\
  }\href@noop {} {\bibfield  {journal} {\bibinfo  {journal} {Phys. Rev. A}\
  }\textbf {\bibinfo {volume} {79}},\ \bibinfo {pages} {023605} (\bibinfo
  {year} {2009})}\BibitemShut {NoStop}%
\bibitem [{\citenamefont {Bulgac}\ and\ \citenamefont
  {Forbes}(2013)}]{bulgac2013TSLDA}%
  \BibitemOpen
  \bibfield  {author} {\bibinfo {author} {\bibfnamefont {A.}~\bibnamefont
  {Bulgac}}\ and\ \bibinfo {author} {\bibfnamefont {M.~M.}\ \bibnamefont
  {Forbes}},\ }\href@noop {} {\emph {\bibinfo {title} {Quantum Gases: Finite
  Temperature and Non-Equilibrium Dynamics (Vol. 1 Cold Atoms Series).}}},\
  edited by\ \bibinfo {editor} {\bibfnamefont {M.~D.}\ \bibnamefont
  {N.P.~Proukakis}, \bibfnamefont {S.A.~Gardiner}}\ and\ \bibinfo {editor}
  {\bibfnamefont {M.}~\bibnamefont {Szymanska}}\ (\bibinfo  {publisher}
  {Imperial College Press},\ \bibinfo {address} {London},\ \bibinfo {year}
  {2013})\BibitemShut {NoStop}%
\bibitem [{\citenamefont {Zwerger}(2011)}]{Zwerger2011BECBCS}%
  \BibitemOpen
  \bibinfo {editor} {\bibfnamefont {W.}~\bibnamefont {Zwerger}},\ ed.,\
  \href@noop {} {\emph {\bibinfo {title} {The BCS-BEC crossover and the unitary
  {Fermi} gas}}},\ Vol.\ \bibinfo {volume} {836}\ (\bibinfo  {publisher}
  {Springer},\ \bibinfo {year} {2011})\BibitemShut {NoStop}%
\bibitem [{\citenamefont {Muryshev}\ \emph {et~al.}(1999)\citenamefont
  {Muryshev}, \citenamefont {van Linden van~den Heuvell},\ and\ \citenamefont
  {Shlyapnikov}}]{Muryshev1999Soliton}%
  \BibitemOpen
  \bibfield  {author} {\bibinfo {author} {\bibfnamefont {A.~E.}\ \bibnamefont
  {Muryshev}}, \bibinfo {author} {\bibfnamefont {H.~B.}\ \bibnamefont {van
  Linden van~den Heuvell}}, \ and\ \bibinfo {author} {\bibfnamefont {G.~V.}\
  \bibnamefont {Shlyapnikov}},\ }\href@noop {} {\bibfield  {journal} {\bibinfo
  {journal} {Phys. Rev. A}\ }\textbf {\bibinfo {volume} {60}},\ \bibinfo
  {pages} {R2665} (\bibinfo {year} {1999})}\BibitemShut {NoStop}%
\bibitem [{\citenamefont {Feder}\ \emph {et~al.}(2000)\citenamefont {Feder},
  \citenamefont {Pindzola}, \citenamefont {Collins}, \citenamefont
  {Schneider},\ and\ \citenamefont {Clark}}]{Feder2000Soliton}%
  \BibitemOpen
  \bibfield  {author} {\bibinfo {author} {\bibfnamefont {D.~L.}\ \bibnamefont
  {Feder}}, \bibinfo {author} {\bibfnamefont {M.~S.}\ \bibnamefont {Pindzola}},
  \bibinfo {author} {\bibfnamefont {L.~A.}\ \bibnamefont {Collins}}, \bibinfo
  {author} {\bibfnamefont {B.~I.}\ \bibnamefont {Schneider}}, \ and\ \bibinfo
  {author} {\bibfnamefont {C.~W.}\ \bibnamefont {Clark}},\ }\href@noop {}
  {\bibfield  {journal} {\bibinfo  {journal} {Phys. Rev. A}\ }\textbf {\bibinfo
  {volume} {62}},\ \bibinfo {pages} {053606} (\bibinfo {year}
  {2000})}\BibitemShut {NoStop}%
\bibitem [{\citenamefont {Brand}\ and\ \citenamefont
  {Reinhardt}(2001)}]{brand01svort}%
  \BibitemOpen
  \bibfield  {author} {\bibinfo {author} {\bibfnamefont {J.}~\bibnamefont
  {Brand}}\ and\ \bibinfo {author} {\bibfnamefont {W.}~\bibnamefont
  {Reinhardt}},\ }\href@noop {} {\bibfield  {journal} {\bibinfo  {journal} {J.
  Phys. B: At. Mol. Opt. Phys.}\ }\textbf {\bibinfo {volume} {34}},\ \bibinfo
  {pages} {L113} (\bibinfo {year} {2001})}\BibitemShut {NoStop}%
\bibitem [{\citenamefont {Brand}\ and\ \citenamefont
  {Reinhardt}(2002)}]{Brand2002solitonicvortex}%
  \BibitemOpen
  \bibfield  {author} {\bibinfo {author} {\bibfnamefont {J.}~\bibnamefont
  {Brand}}\ and\ \bibinfo {author} {\bibfnamefont {W.~P.}\ \bibnamefont
  {Reinhardt}},\ }\href@noop {} {\bibfield  {journal} {\bibinfo  {journal}
  {Phys. Rev. A}\ }\textbf {\bibinfo {volume} {65}},\ \bibinfo {pages} {043612}
  (\bibinfo {year} {2002})}\BibitemShut {NoStop}%
\bibitem [{\citenamefont {Komineas}\ and\ \citenamefont
  {Papanicolaou}(2003)}]{Komineas2003soliton}%
  \BibitemOpen
  \bibfield  {author} {\bibinfo {author} {\bibfnamefont {S.}~\bibnamefont
  {Komineas}}\ and\ \bibinfo {author} {\bibfnamefont {N.}~\bibnamefont
  {Papanicolaou}},\ }\href@noop {} {\bibfield  {journal} {\bibinfo  {journal}
  {Phys. Rev. A}\ }\textbf {\bibinfo {volume} {68}},\ \bibinfo {pages} {043617}
  (\bibinfo {year} {2003})}\BibitemShut {NoStop}%
\bibitem [{\citenamefont {Mu\~noz Mateo}\ and\ \citenamefont
  {Brand}(2014)}]{mateo2014chladni}%
  \BibitemOpen
  \bibfield  {author} {\bibinfo {author} {\bibfnamefont {A.}~\bibnamefont
  {Mu\~noz Mateo}}\ and\ \bibinfo {author} {\bibfnamefont {J.}~\bibnamefont
  {Brand}},\ }\href {\doibase 10.1103/PhysRevLett.113.255302} {\bibfield
  {journal} {\bibinfo  {journal} {Phys. Rev. Lett.}\ }\textbf {\bibinfo
  {volume} {113}},\ \bibinfo {pages} {255302} (\bibinfo {year}
  {2014})}\BibitemShut {NoStop}%
\bibitem [{\citenamefont {Anderson}\ \emph {et~al.}(2001)\citenamefont
  {Anderson}, \citenamefont {Haljan}, \citenamefont {Regal}, \citenamefont
  {Feder}, \citenamefont {Collins}, \citenamefont {Clark},\ and\ \citenamefont
  {Cornell}}]{ande01ring}%
  \BibitemOpen
  \bibfield  {author} {\bibinfo {author} {\bibfnamefont {B.~P.}\ \bibnamefont
  {Anderson}}, \bibinfo {author} {\bibfnamefont {P.~C.}\ \bibnamefont
  {Haljan}}, \bibinfo {author} {\bibfnamefont {C.~A.}\ \bibnamefont {Regal}},
  \bibinfo {author} {\bibfnamefont {D.~L.}\ \bibnamefont {Feder}}, \bibinfo
  {author} {\bibfnamefont {L.~A.}\ \bibnamefont {Collins}}, \bibinfo {author}
  {\bibfnamefont {C.~W.}\ \bibnamefont {Clark}}, \ and\ \bibinfo {author}
  {\bibfnamefont {E.~A.}\ \bibnamefont {Cornell}},\ }\href@noop {} {\bibfield
  {journal} {\bibinfo  {journal} {Phys. Rev. Lett.}\ }\textbf {\bibinfo
  {volume} {86}},\ \bibinfo {pages} {2926} (\bibinfo {year}
  {2001})}\BibitemShut {NoStop}%
\bibitem [{\citenamefont {Dutton}\ \emph {et~al.}(2001)\citenamefont {Dutton},
  \citenamefont {Budde}, \citenamefont {Slowe},\ and\ \citenamefont
  {Hau}}]{dutton2001shock}%
  \BibitemOpen
  \bibfield  {author} {\bibinfo {author} {\bibfnamefont {Z.}~\bibnamefont
  {Dutton}}, \bibinfo {author} {\bibfnamefont {M.}~\bibnamefont {Budde}},
  \bibinfo {author} {\bibfnamefont {C.}~\bibnamefont {Slowe}}, \ and\ \bibinfo
  {author} {\bibfnamefont {L.~V.}\ \bibnamefont {Hau}},\ }\href@noop {}
  {\bibfield  {journal} {\bibinfo  {journal} {Science}\ }\textbf {\bibinfo
  {volume} {293}},\ \bibinfo {pages} {663} (\bibinfo {year}
  {2001})}\BibitemShut {NoStop}%
\bibitem [{\citenamefont {Becker}\ \emph {et~al.}(2013)\citenamefont {Becker},
  \citenamefont {Sengstock}, \citenamefont {Schmelcher}, \citenamefont
  {Kevrekidis},\ and\ \citenamefont
  {Carretero-GonzÃ¡lez}}]{Becker2013SolitonicVortex}%
  \BibitemOpen
  \bibfield  {author} {\bibinfo {author} {\bibfnamefont {C.}~\bibnamefont
  {Becker}}, \bibinfo {author} {\bibfnamefont {K.}~\bibnamefont {Sengstock}},
  \bibinfo {author} {\bibfnamefont {P.}~\bibnamefont {Schmelcher}}, \bibinfo
  {author} {\bibfnamefont {P.~G.}\ \bibnamefont {Kevrekidis}}, \ and\ \bibinfo
  {author} {\bibfnamefont {R.}~\bibnamefont {Carretero-GonzÃ¡lez}},\
  }\href@noop {} {\bibfield  {journal} {\bibinfo  {journal} {New Journal of
  Physics}\ }\textbf {\bibinfo {volume} {15}},\ \bibinfo {pages} {113028}
  (\bibinfo {year} {2013})}\BibitemShut {NoStop}%
\bibitem [{\citenamefont {Wlaz\l{}owski}\ \emph {et~al.}(2015)\citenamefont
  {Wlaz\l{}owski}, \citenamefont {Bulgac}, \citenamefont {Forbes},\ and\
  \citenamefont {Roche}}]{wlaz14lifecycle}%
  \BibitemOpen
  \bibfield  {author} {\bibinfo {author} {\bibfnamefont {G.}~\bibnamefont
  {Wlaz\l{}owski}}, \bibinfo {author} {\bibfnamefont {A.}~\bibnamefont
  {Bulgac}}, \bibinfo {author} {\bibfnamefont {M.~M.}\ \bibnamefont {Forbes}},
  \ and\ \bibinfo {author} {\bibfnamefont {K.~J.}\ \bibnamefont {Roche}},\
  }\href {\doibase 10.1103/PhysRevA.91.031602} {\bibfield  {journal} {\bibinfo
  {journal} {Phys. Rev. A}\ }\textbf {\bibinfo {volume} {91}},\ \bibinfo
  {pages} {031602} (\bibinfo {year} {2015})}\BibitemShut {NoStop}%
\bibitem [{\citenamefont {Reichl}\ and\ \citenamefont
  {Mueller}(2013)}]{Reichl2013VortexRing}%
  \BibitemOpen
  \bibfield  {author} {\bibinfo {author} {\bibfnamefont {M.~D.}\ \bibnamefont
  {Reichl}}\ and\ \bibinfo {author} {\bibfnamefont {E.~J.}\ \bibnamefont
  {Mueller}},\ }\href@noop {} {\bibfield  {journal} {\bibinfo  {journal} {Phys.
  Rev. A}\ }\textbf {\bibinfo {volume} {88}},\ \bibinfo {pages} {053626}
  (\bibinfo {year} {2013})}\BibitemShut {NoStop}%
\bibitem [{\citenamefont {Scherpelz}\ \emph {et~al.}(2014)\citenamefont
  {Scherpelz}, \citenamefont {Padavić}, \citenamefont {Rançon}, \citenamefont
  {Glatz}, \citenamefont {Aranson},\ and\ \citenamefont
  {Levin}}]{scherpelz2014vortexring}%
  \BibitemOpen
  \bibfield  {author} {\bibinfo {author} {\bibfnamefont {P.}~\bibnamefont
  {Scherpelz}}, \bibinfo {author} {\bibfnamefont {K.}~\bibnamefont {Padavić}},
  \bibinfo {author} {\bibfnamefont {A.}~\bibnamefont {Rançon}}, \bibinfo
  {author} {\bibfnamefont {A.}~\bibnamefont {Glatz}}, \bibinfo {author}
  {\bibfnamefont {I.~S.}\ \bibnamefont {Aranson}}, \ and\ \bibinfo {author}
  {\bibfnamefont {K.}~\bibnamefont {Levin}},\ }\href@noop {} {\bibfield
  {journal} {\bibinfo  {journal} {Phys. Rev. Lett.}\ }\textbf {\bibinfo
  {volume} {113}},\ \bibinfo {pages} {125301} (\bibinfo {year}
  {2014})}\BibitemShut {NoStop}%
\bibitem [{\citenamefont {Ku}\ \emph {et~al.}(2014)\citenamefont {Ku},
  \citenamefont {Ji}, \citenamefont {Mukherjee}, \citenamefont {Sanchez},
  \citenamefont {Cheuk}, \citenamefont {Yefsah},\ and\ \citenamefont
  {Zwierlein}}]{Ku2014svortex}%
  \BibitemOpen
  \bibfield  {author} {\bibinfo {author} {\bibfnamefont {M.~J.-H.}\
  \bibnamefont {Ku}}, \bibinfo {author} {\bibfnamefont {W.}~\bibnamefont {Ji}},
  \bibinfo {author} {\bibfnamefont {B.}~\bibnamefont {Mukherjee}}, \bibinfo
  {author} {\bibfnamefont {E.~G.}\ \bibnamefont {Sanchez}}, \bibinfo {author}
  {\bibfnamefont {L.~T.}\ \bibnamefont {Cheuk}}, \bibinfo {author}
  {\bibfnamefont {T.}~\bibnamefont {Yefsah}}, \ and\ \bibinfo {author}
  {\bibfnamefont {M.~W.}\ \bibnamefont {Zwierlein}},\ }\href@noop {} {\bibfield
   {journal} {\bibinfo  {journal} {Phys. Rev. Lett.}\ }\textbf {\bibinfo
  {volume} {113}},\ \bibinfo {pages} {065301} (\bibinfo {year}
  {2014})}\BibitemShut {NoStop}%
\bibitem [{\citenamefont {Ketterle}\ and\ \citenamefont
  {Zwierlein}(2008{\natexlab{a}})}]{kett08rivista}%
  \BibitemOpen
  \bibfield  {author} {\bibinfo {author} {\bibfnamefont {W.}~\bibnamefont
  {Ketterle}}\ and\ \bibinfo {author} {\bibfnamefont {M.}~\bibnamefont
  {Zwierlein}},\ }\href@noop {} {\bibfield  {journal} {\bibinfo  {journal}
  {Rivista del Nuovo Cimento}\ }\textbf {\bibinfo {volume} {31}},\ \bibinfo
  {pages} {247} (\bibinfo {year} {2008}{\natexlab{a}})}\BibitemShut {NoStop}%
\bibitem [{see()}]{seeSuppmat}%
  \BibitemOpen
  \href@noop {} {\bibinfo  {journal} {See Supplemental Material}\ }\BibitemShut
  {NoStop}%
\bibitem [{\citenamefont {Burger}\ \emph {et~al.}(1999)\citenamefont {Burger},
  \citenamefont {Bongs}, \citenamefont {Dettmer}, \citenamefont {Ertmer},\ and\
  \citenamefont {Sengstock}}]{burg99soliton}%
  \BibitemOpen
\bibfield  {journal} {  }\bibfield  {author} {\bibinfo {author} {\bibfnamefont
  {S.}~\bibnamefont {Burger}}, \bibinfo {author} {\bibfnamefont
  {K.}~\bibnamefont {Bongs}}, \bibinfo {author} {\bibfnamefont
  {S.}~\bibnamefont {Dettmer}}, \bibinfo {author} {\bibfnamefont
  {W.}~\bibnamefont {Ertmer}}, \ and\ \bibinfo {author} {\bibfnamefont
  {K.}~\bibnamefont {Sengstock}},\ }\href@noop {} {\bibfield  {journal}
  {\bibinfo  {journal} {Phys. Rev. Lett.}\ }\textbf {\bibinfo {volume} {83}},\
  \bibinfo {pages} {5198} (\bibinfo {year} {1999})}\BibitemShut {NoStop}%
\bibitem [{\citenamefont {Denschlag}\ \emph {et~al.}(2000)\citenamefont
  {Denschlag}, \citenamefont {Simsarian}, \citenamefont {Feder}, \citenamefont
  {Clark}, \citenamefont {Collins}, \citenamefont {Cubizolles}, \citenamefont
  {Deng}, \citenamefont {Hagley}, \citenamefont {Helmerson}, \citenamefont
  {Reinhardt}, \citenamefont {Rolston}, \citenamefont {Schneider},\ and\
  \citenamefont {Phillips}}]{dens00}%
  \BibitemOpen
  \bibfield  {author} {\bibinfo {author} {\bibfnamefont {J.}~\bibnamefont
  {Denschlag}}, \bibinfo {author} {\bibfnamefont {J.~E.}\ \bibnamefont
  {Simsarian}}, \bibinfo {author} {\bibfnamefont {D.~L.}\ \bibnamefont
  {Feder}}, \bibinfo {author} {\bibfnamefont {C.~W.}\ \bibnamefont {Clark}},
  \bibinfo {author} {\bibfnamefont {L.~A.}\ \bibnamefont {Collins}}, \bibinfo
  {author} {\bibfnamefont {J.}~\bibnamefont {Cubizolles}}, \bibinfo {author}
  {\bibfnamefont {L.}~\bibnamefont {Deng}}, \bibinfo {author} {\bibfnamefont
  {E.~W.}\ \bibnamefont {Hagley}}, \bibinfo {author} {\bibfnamefont
  {K.}~\bibnamefont {Helmerson}}, \bibinfo {author} {\bibfnamefont {W.~P.}\
  \bibnamefont {Reinhardt}}, \bibinfo {author} {\bibfnamefont {S.~L.}\
  \bibnamefont {Rolston}}, \bibinfo {author} {\bibfnamefont {B.~I.}\
  \bibnamefont {Schneider}}, \ and\ \bibinfo {author} {\bibfnamefont {W.~D.}\
  \bibnamefont {Phillips}},\ }\href@noop {} {\bibfield  {journal} {\bibinfo
  {journal} {Science}\ }\textbf {\bibinfo {volume} {287}},\ \bibinfo {pages}
  {97} (\bibinfo {year} {2000})}\BibitemShut {NoStop}%
\bibitem [{\citenamefont {Becker}\ \emph {et~al.}(2008)\citenamefont {Becker},
  \citenamefont {Stellmer}, \citenamefont {Soltan-Panahi}, \citenamefont
  {D{\"o}scher}, \citenamefont {Baumert}, \citenamefont {Richter},
  \citenamefont {Kronj{\"a}ger}, \citenamefont {Bongs},\ and\ \citenamefont
  {Sengstock}}]{Becker2008solitons}%
  \BibitemOpen
  \bibfield  {author} {\bibinfo {author} {\bibfnamefont {C.}~\bibnamefont
  {Becker}}, \bibinfo {author} {\bibfnamefont {S.}~\bibnamefont {Stellmer}},
  \bibinfo {author} {\bibfnamefont {P.}~\bibnamefont {Soltan-Panahi}}, \bibinfo
  {author} {\bibfnamefont {S.}~\bibnamefont {D{\"o}scher}}, \bibinfo {author}
  {\bibfnamefont {M.}~\bibnamefont {Baumert}}, \bibinfo {author} {\bibfnamefont
  {E.~M.}\ \bibnamefont {Richter}}, \bibinfo {author} {\bibfnamefont
  {J.}~\bibnamefont {Kronj{\"a}ger}}, \bibinfo {author} {\bibfnamefont
  {K.}~\bibnamefont {Bongs}}, \ and\ \bibinfo {author} {\bibfnamefont
  {K.}~\bibnamefont {Sengstock}},\ }\href@noop {} {\bibfield  {journal}
  {\bibinfo  {journal} {Nature Physics}\ }\textbf {\bibinfo {volume} {4}},\
  \bibinfo {pages} {496} (\bibinfo {year} {2008})}\BibitemShut {NoStop}%
\bibitem [{\citenamefont {Yefsah}\ \emph {et~al.}(2013)\citenamefont {Yefsah},
  \citenamefont {Sommer}, \citenamefont {Ku}, \citenamefont {Cheuk},
  \citenamefont {Ji}, \citenamefont {Bakr},\ and\ \citenamefont
  {Zwierlein}}]{Yefsah2013Soliton}%
  \BibitemOpen
  \bibfield  {author} {\bibinfo {author} {\bibfnamefont {T.}~\bibnamefont
  {Yefsah}}, \bibinfo {author} {\bibfnamefont {A.~T.}\ \bibnamefont {Sommer}},
  \bibinfo {author} {\bibfnamefont {M.~J.~H.}\ \bibnamefont {Ku}}, \bibinfo
  {author} {\bibfnamefont {L.~W.}\ \bibnamefont {Cheuk}}, \bibinfo {author}
  {\bibfnamefont {W.}~\bibnamefont {Ji}}, \bibinfo {author} {\bibfnamefont
  {W.~S.}\ \bibnamefont {Bakr}}, \ and\ \bibinfo {author} {\bibfnamefont
  {M.~W.}\ \bibnamefont {Zwierlein}},\ }\href@noop {} {\bibfield  {journal}
  {\bibinfo  {journal} {Nature}\ }\textbf {\bibinfo {volume} {499}},\ \bibinfo
  {pages} {426} (\bibinfo {year} {2013})}\BibitemShut {NoStop}%
\bibitem [{\citenamefont {Ketterle}\ and\ \citenamefont
  {Zwierlein}(2008{\natexlab{b}})}]{kett08varenna}%
  \BibitemOpen
  \bibfield  {author} {\bibinfo {author} {\bibfnamefont {W.}~\bibnamefont
  {Ketterle}}\ and\ \bibinfo {author} {\bibfnamefont {M.}~\bibnamefont
  {Zwierlein}},\ }in\ \href@noop {} {\emph {\bibinfo {booktitle} {Ultracold
  Fermi Gases, Proceedings of the International School of Physics "Enrico
  Fermi", Course CLXIV, Varenna, 20 - 30 June 2006.}}},\ \bibinfo {editor}
  {edited by\ \bibinfo {editor} {\bibfnamefont {M.}~\bibnamefont {Inguscio}},
  \bibinfo {editor} {\bibfnamefont {W.}~\bibnamefont {Ketterle}}, \ and\
  \bibinfo {editor} {\bibfnamefont {C.}~\bibnamefont {Salomon}}}\ (\bibinfo
  {publisher} {IOS Press},\ \bibinfo {address} {Amsterdam.},\ \bibinfo {year}
  {2008})\BibitemShut {NoStop}%
\bibitem [{\citenamefont {Zwierlein}\ \emph {et~al.}(2005)\citenamefont
  {Zwierlein}, \citenamefont {Abo-Shaeer}, \citenamefont {Schirotzek},
  \citenamefont {Schunck},\ and\ \citenamefont {Ketterle}}]{zwie05vort}%
  \BibitemOpen
  \bibfield  {author} {\bibinfo {author} {\bibfnamefont {M.~W.}\ \bibnamefont
  {Zwierlein}}, \bibinfo {author} {\bibfnamefont {J.~R.}\ \bibnamefont
  {Abo-Shaeer}}, \bibinfo {author} {\bibfnamefont {A.}~\bibnamefont
  {Schirotzek}}, \bibinfo {author} {\bibfnamefont {C.~H.}\ \bibnamefont
  {Schunck}}, \ and\ \bibinfo {author} {\bibfnamefont {W.}~\bibnamefont
  {Ketterle}},\ }\href@noop {} {\bibfield  {journal} {\bibinfo  {journal}
  {Nature}\ }\textbf {\bibinfo {volume} {435}},\ \bibinfo {pages} {1047}
  (\bibinfo {year} {2005})}\BibitemShut {NoStop}%
\bibitem [{\citenamefont {Ku}\ \emph {et~al.}(2012)\citenamefont {Ku},
  \citenamefont {Sommer}, \citenamefont {Cheuk},\ and\ \citenamefont
  {Zwierlein}}]{ku2012thermodynamics}%
  \BibitemOpen
  \bibfield  {author} {\bibinfo {author} {\bibfnamefont {M.~J.~H.}\
  \bibnamefont {Ku}}, \bibinfo {author} {\bibfnamefont {A.~T.}\ \bibnamefont
  {Sommer}}, \bibinfo {author} {\bibfnamefont {L.~W.}\ \bibnamefont {Cheuk}}, \
  and\ \bibinfo {author} {\bibfnamefont {M.~W.}\ \bibnamefont {Zwierlein}},\
  }\href@noop {} {\bibfield  {journal} {\bibinfo  {journal} {Science}\ }\textbf
  {\bibinfo {volume} {335}},\ \bibinfo {pages} {563} (\bibinfo {year}
  {2012})}\BibitemShut {NoStop}%
\bibitem [{Note1()}]{Note1}%
  \BibitemOpen
  \bibinfo {note} {The observations made here are reminiscent of the first
  phase imprinting experiment realized in Bose-Einstein condensates~\cite
  {burg99soliton}, where two waves rapidly propagating in opposite directions
  were created, one of which has been interpreted as a fast dark soliton. Here,
  we observe that both wavefronts propagate at the speed of sound.}\BibitemShut
  {Stop}%
\bibitem [{\citenamefont {Sacha}\ and\ \citenamefont
  {Delande}(2014)}]{sacha14imprint}%
  \BibitemOpen
  \bibfield  {author} {\bibinfo {author} {\bibfnamefont {K.}~\bibnamefont
  {Sacha}}\ and\ \bibinfo {author} {\bibfnamefont {D.}~\bibnamefont
  {Delande}},\ }\href {\doibase 10.1103/PhysRevA.90.021604} {\bibfield
  {journal} {\bibinfo  {journal} {Phys. Rev. A}\ }\textbf {\bibinfo {volume}
  {90}},\ \bibinfo {pages} {021604} (\bibinfo {year} {2014})}\BibitemShut
  {NoStop}%
\bibitem [{\citenamefont {{Scherpelz}}\ \emph {et~al.}()\citenamefont
  {{Scherpelz}}, \citenamefont {{Padavi{\'c}}}, \citenamefont {{Murray}},
  \citenamefont {{Glatz}}, \citenamefont {{Aranson}},\ and\ \citenamefont
  {{Levin}}}]{Scherpelz2014defects}%
  \BibitemOpen
  \bibfield  {author} {\bibinfo {author} {\bibfnamefont {P.}~\bibnamefont
  {{Scherpelz}}}, \bibinfo {author} {\bibfnamefont {K.}~\bibnamefont
  {{Padavi{\'c}}}}, \bibinfo {author} {\bibfnamefont {A.}~\bibnamefont
  {{Murray}}}, \bibinfo {author} {\bibfnamefont {A.}~\bibnamefont {{Glatz}}},
  \bibinfo {author} {\bibfnamefont {I.~S.}\ \bibnamefont {{Aranson}}}, \ and\
  \bibinfo {author} {\bibfnamefont {K.}~\bibnamefont {{Levin}}},\ }\href@noop
  {} {\bibfield  {journal} {\bibinfo  {journal} {ArXiv e-prints}\ }}\Eprint
  {http://arxiv.org/abs/1410.0067} {1410.0067} \BibitemShut {NoStop}%
\bibitem [{\citenamefont {Bulgac}\ \emph {et~al.}(2014)\citenamefont {Bulgac},
  \citenamefont {Forbes}, \citenamefont {Kelley}, \citenamefont {Roche},\ and\
  \citenamefont {Wlazlowski}}]{Bulgac2014Vortexrings}%
  \BibitemOpen
  \bibfield  {author} {\bibinfo {author} {\bibfnamefont {A.}~\bibnamefont
  {Bulgac}}, \bibinfo {author} {\bibfnamefont {M.~M.}\ \bibnamefont {Forbes}},
  \bibinfo {author} {\bibfnamefont {M.~M.}\ \bibnamefont {Kelley}}, \bibinfo
  {author} {\bibfnamefont {K.~J.}\ \bibnamefont {Roche}}, \ and\ \bibinfo
  {author} {\bibfnamefont {G.}~\bibnamefont {Wlazlowski}},\ }\href@noop {}
  {\bibfield  {journal} {\bibinfo  {journal} {Phys. Rev. Lett.}\ }\textbf
  {\bibinfo {volume} {112}},\ \bibinfo {pages} {025301} (\bibinfo {year}
  {2014})}\BibitemShut {NoStop}%
\bibitem [{\citenamefont {Wen}\ \emph {et~al.}(2013)\citenamefont {Wen},
  \citenamefont {Zhao},\ and\ \citenamefont {Ma}}]{Wen2013DarkSoliton}%
  \BibitemOpen
  \bibfield  {author} {\bibinfo {author} {\bibfnamefont {W.}~\bibnamefont
  {Wen}}, \bibinfo {author} {\bibfnamefont {C.}~\bibnamefont {Zhao}}, \ and\
  \bibinfo {author} {\bibfnamefont {X.}~\bibnamefont {Ma}},\ }\href@noop {}
  {\bibfield  {journal} {\bibinfo  {journal} {Phys. Rev. A}\ }\textbf {\bibinfo
  {volume} {88}},\ \bibinfo {pages} {063621} (\bibinfo {year}
  {2013})}\BibitemShut {NoStop}%
\bibitem [{\citenamefont {Brand}()}]{Brandprivatecommunication}%
  \BibitemOpen
  \bibfield  {author} {\bibinfo {author} {\bibfnamefont {J.}~\bibnamefont
  {Brand}},\ }\href@noop {} {\bibinfo  {journal} {Private Communication}\
  }\BibitemShut {NoStop}%
\bibitem [{\citenamefont {Kamchatnov}\ and\ \citenamefont
  {Pitaevskii}(2008)}]{Kamchatnov2008}%
  \BibitemOpen
\bibfield  {journal} {  }\bibfield  {author} {\bibinfo {author} {\bibfnamefont
  {A.~M.}\ \bibnamefont {Kamchatnov}}\ and\ \bibinfo {author} {\bibfnamefont
  {L.~P.}\ \bibnamefont {Pitaevskii}},\ }\href {\doibase
  10.1103/PhysRevLett.100.160402} {\bibfield  {journal} {\bibinfo  {journal}
  {Phys. Rev. Lett.}\ }\textbf {\bibinfo {volume} {100}},\ \bibinfo {pages}
  {160402} (\bibinfo {year} {2008})}\BibitemShut {NoStop}%
\bibitem [{Note2()}]{Note2}%
  \BibitemOpen
  \bibinfo {note} {Note that the growth rate that we report here is in the
  units of a speed, and not a frequency as considered in~\cite
  {Kamchatnov2008,scott2012solitondecay}.}\BibitemShut {Stop}%
\bibitem [{\citenamefont {Scott}\ \emph {et~al.}(2012)\citenamefont {Scott},
  \citenamefont {Dalfovo}, \citenamefont {Pitaevskii}, \citenamefont
  {Stringari}, \citenamefont {Fialko}, \citenamefont {Liao},\ and\
  \citenamefont {Brand}}]{scott2012solitondecay}%
  \BibitemOpen
  \bibfield  {author} {\bibinfo {author} {\bibfnamefont {R.~G.}\ \bibnamefont
  {Scott}}, \bibinfo {author} {\bibfnamefont {F.}~\bibnamefont {Dalfovo}},
  \bibinfo {author} {\bibfnamefont {L.~P.}\ \bibnamefont {Pitaevskii}},
  \bibinfo {author} {\bibfnamefont {S.}~\bibnamefont {Stringari}}, \bibinfo
  {author} {\bibfnamefont {O.}~\bibnamefont {Fialko}}, \bibinfo {author}
  {\bibfnamefont {R.}~\bibnamefont {Liao}}, \ and\ \bibinfo {author}
  {\bibfnamefont {J.}~\bibnamefont {Brand}},\ }\href@noop {} {\bibfield
  {journal} {\bibinfo  {journal} {New Journal of Physics}\ }\textbf {\bibinfo
  {volume} {14}},\ \bibinfo {pages} {023044} (\bibinfo {year}
  {2012})}\BibitemShut {NoStop}%
\bibitem [{\citenamefont {Spuntarelli}\ \emph {et~al.}(2011)\citenamefont
  {Spuntarelli}, \citenamefont {Carr}, \citenamefont {Pieri},\ and\
  \citenamefont {Strinati}}]{spuntarelli2011soliton}%
  \BibitemOpen
  \bibfield  {author} {\bibinfo {author} {\bibfnamefont {A.}~\bibnamefont
  {Spuntarelli}}, \bibinfo {author} {\bibfnamefont {L.~D.}\ \bibnamefont
  {Carr}}, \bibinfo {author} {\bibfnamefont {P.}~\bibnamefont {Pieri}}, \ and\
  \bibinfo {author} {\bibfnamefont {G.~C.}\ \bibnamefont {Strinati}},\
  }\href@noop {} {\bibfield  {journal} {\bibinfo  {journal} {New Journal of
  Physics}\ }\textbf {\bibinfo {volume} {13}},\ \bibinfo {pages} {035010}
  (\bibinfo {year} {2011})}\BibitemShut {NoStop}%
\bibitem [{\citenamefont {Yoshida}\ and\ \citenamefont
  {Yip}(2007)}]{Yoshida2007LO}%
  \BibitemOpen
  \bibfield  {author} {\bibinfo {author} {\bibfnamefont {N.}~\bibnamefont
  {Yoshida}}\ and\ \bibinfo {author} {\bibfnamefont {S.~K.}\ \bibnamefont
  {Yip}},\ }\href@noop {} {\bibfield  {journal} {\bibinfo  {journal} {Phys.
  Rev. A}\ }\textbf {\bibinfo {volume} {75}},\ \bibinfo {pages} {063601}
  (\bibinfo {year} {2007})}\BibitemShut {NoStop}%
\bibitem [{\citenamefont {Radzihovsky}(2011)}]{Radz2011LO}%
  \BibitemOpen
  \bibfield  {author} {\bibinfo {author} {\bibfnamefont {L.}~\bibnamefont
  {Radzihovsky}},\ }\href@noop {} {\bibfield  {journal} {\bibinfo  {journal}
  {Phys. Rev. A}\ }\textbf {\bibinfo {volume} {84}},\ \bibinfo {pages} {023611}
  (\bibinfo {year} {2011})}\BibitemShut {NoStop}%
\bibitem [{\citenamefont {Lutchyn}\ \emph {et~al.}(2011)\citenamefont
  {Lutchyn}, \citenamefont {Dzero},\ and\ \citenamefont
  {Yakovenko}}]{Lutchyn2011Soliton}%
  \BibitemOpen
  \bibfield  {author} {\bibinfo {author} {\bibfnamefont {R.~M.}\ \bibnamefont
  {Lutchyn}}, \bibinfo {author} {\bibfnamefont {M.}~\bibnamefont {Dzero}}, \
  and\ \bibinfo {author} {\bibfnamefont {V.~M.}\ \bibnamefont {Yakovenko}},\
  }\href@noop {} {\bibfield  {journal} {\bibinfo  {journal} {Phys. Rev. A}\
  }\textbf {\bibinfo {volume} {84}},\ \bibinfo {pages} {033609} (\bibinfo
  {year} {2011})}\BibitemShut {NoStop}%
\end{thebibliography}%

\clearpage
\newpage
%%%%%%%%%%%%%%%%%%%%%%%%%%%%%%% SUPP MAT. %%%%%%%%%%%%%%%%%%%%%%%%%%%%%%%

\section{SUPPLEMENTAL MATERIAL}

\textbf{Supplementary data.} In Figs.~\ref{f:SuppFig_FullMovie}, ~\ref{f:SuppFig_FullSnake} and ~\ref{f:SuppFig_FullTomography} we present supplementary data that could not be included in the main text. Fig.~\ref{f:SuppFig_FullMovie} shows the time evolution of the central slice density, after rapid ramp and time of flight, from $t=0$ to $t=19.5\,$ms in time steps of  0.5\,ms. In Fig.~\ref{f:SuppFig_FullSnake}, we show all the images used in the spectral analysis of the snake instability, along with their detected minima and Fourier reconstruction. In Fig.~\ref{f:SuppFig_FullTomography} we show the full tomography recorded at several instances of the dynamics from $t=3.5\,$ms to $t=1000\,$ms.\\

\begin{figure}[h]
    \begin{center}
    \includegraphics[width=60mm]{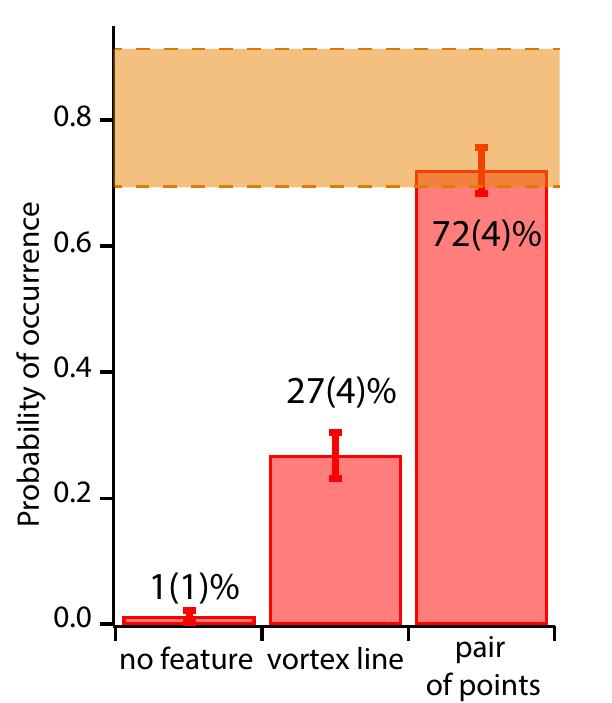}
  \caption{Histogram of events between $t{=}50$\,ms and $t{=}100$\,ms. The probability of detecting no distinctive feature at all is $p_\emptyset{=}1(1)\%$, the probability of detecting a horizontal line is $p=27(4)\%$ and the probability of detecting pairs of nodal points is $p_{\rm{VR}}=72(4)\%$. The shaded orange area is the interval consistent with $3p$ given the uncertainty on $p$.}
    \label{f:stats}
    \end{center}
\end{figure}

\begin{figure*}[h]
    \begin{center}
    \includegraphics[width=175mm]{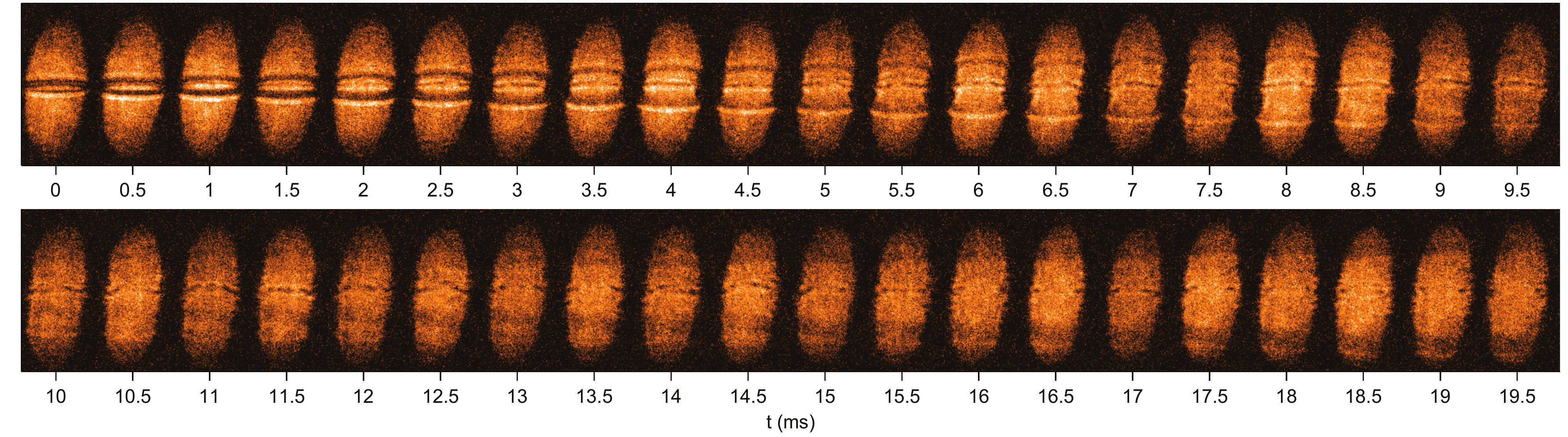}
  \caption{Evolution of the superfluid after the phase imprint, as observed in the density of the central slice ($y=-13\,\mu\mathrm{m}$) of the superfluid, after rapid ramp and time of flight, for $0\,\mathrm{ms}\leq t\leq 19.5\,\mathrm{ms}$, in time steps of 0.5\,ms.}
    \label{f:SuppFig_FullMovie}
    \end{center}
\end{figure*}

\begin{figure*}[h]
    \begin{center}
    \includegraphics[width=175mm]{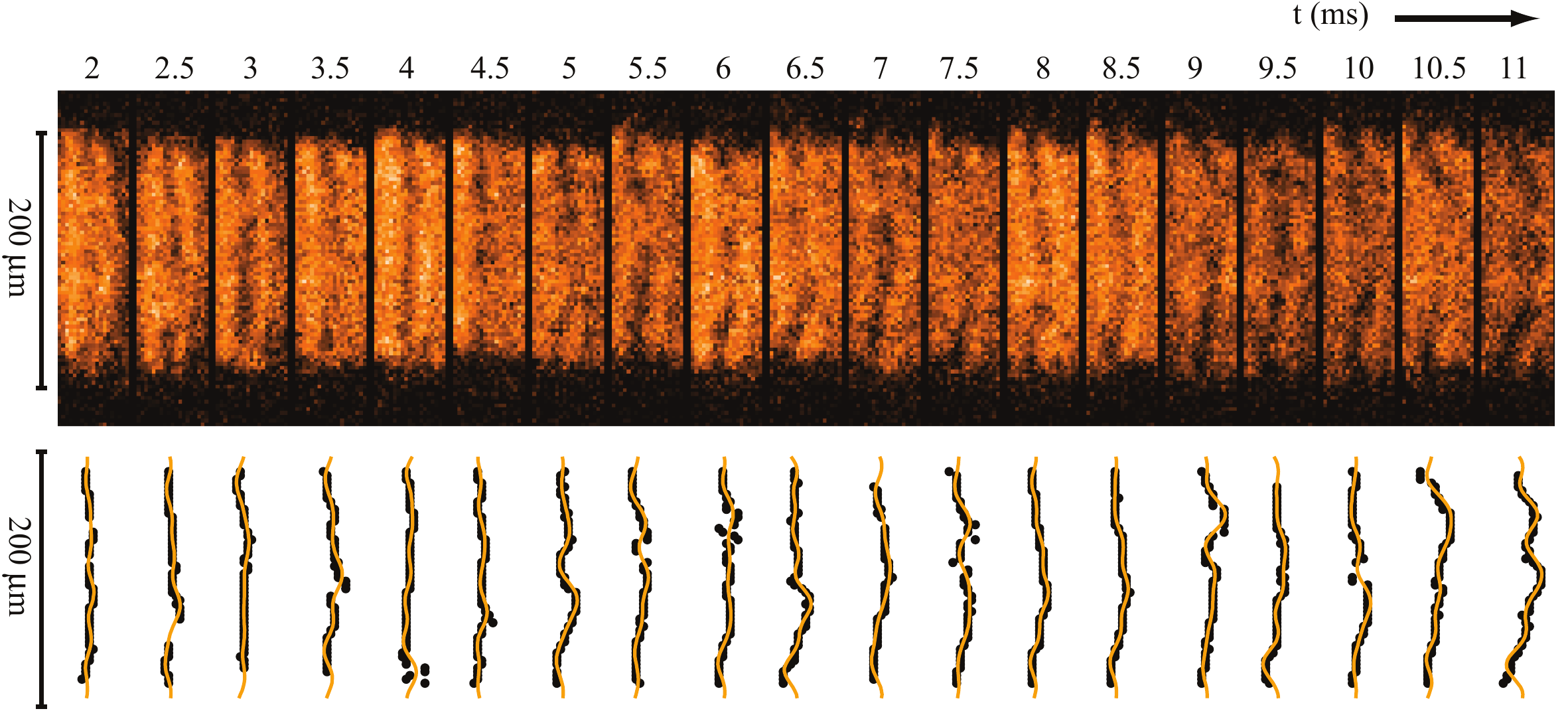}
  \caption{Snake instability of a planar dark soliton in a unitary Fermi gas, observed at early times after the phase imprint. The images represent a time series in time steps of $0.5\,\mathrm{ms}$ after the phase imprint. Top: optical density of the central slice near $z=0$, after rapid ramp and time of flight. Bottom: detected minima (black) and reconstruction using Fourier analysis (gold).}
    \label{f:SuppFig_FullSnake}
    \end{center}
\end{figure*}

\begin{figure*}
    \begin{center}
    \includegraphics[width=120mm]{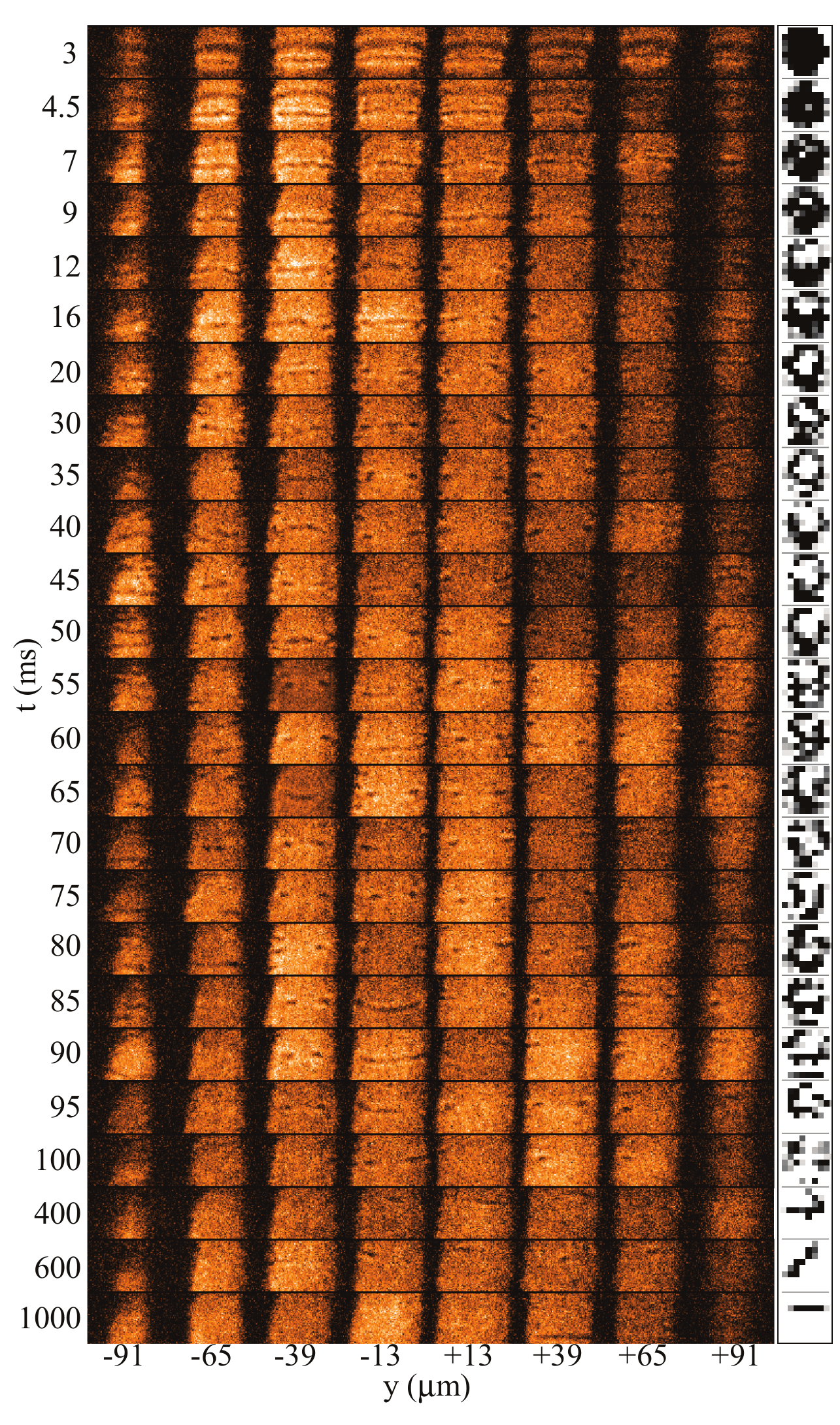}
  \caption{Tomographic images at various times $t$ after the phase-imprint. The right column shows the structure of the solitary waves as if one would observe along the $z$-axis, reconstructed using the residuals of the tomographic images.}
    \label{f:SuppFig_FullTomography}
    \end{center}
\end{figure*}

\textbf{$\Phi$-soliton or not?} The tomographic images taken at the instants $t\sim 50-100$\,ms show both the presence of horizontal vortex lines and pairs of nodal points. Because of the destructive nature of our measurement, one could a priori attribute this observation to two distinct situations: (A) the imprinting leads to either a vortex ring \textit{\textbf{or}} a vortex line with unknown probabilities; (B) the imprinting systematically leads to the formation of a $\Phi$-soliton, \textit{i.e} each shot produces a vortex ring \textit{\textbf{and}} a vortex line. To lift this ambiguity, we establish, for each scenario, the constraints on the probabilities of (i) detecting no depletion (ii) detecting a horizontal vortex line (iii) detecting two nodal dots. Here, we focus on the four central slices near $y=0$ (where horizontal vortex lines are observed). In the case of scenario (A): if $p$ is the probability to find a vortex line in one of the four slices, then the probability of detecting no depletion at all should be $\geq3p$. Allowing for an inequality here accounts for the fact that one cannot exclude shots where the imprinting failed. This relation is true regardless of the probability of having a vortex ring. On the other hand, in the case of scenario (B): if we call $p$ the probability of finding a vortex line in one of the four slices, then the probability of detecting two nodal points should be $3p$. A statistical analysis over $\sim90$ slices recorded between $t=50$\,ms and $t=100$\,ms leads to the histogram shown in Fig.~\ref{f:stats}. The probability of finding a line in one of the four central slices in this time interval is $p=27(4)\%$ while the probabilities of detecting no depletion and a pair of nodal points are respectively $p_\emptyset=1(1)\%$ and $p_{\rm{VR}}=72(4)\%$. This result excludes scenario (A) and strongly supports scenario (B): the systematic formation of a $\Phi$-soliton via a double puncture of the soliton plane.

\textit{Generation of the histogram.} The statistical sample consists of 22 slices at $y=-39\,\mu$m, 30 slices at $y=-13\,\mu$m, 24 slices at $y=13\,\mu$m and 13 slices at $y=39\,\mu$m. The probability is obtained from several runs of random sampling of the data. In a given run, we randomly pick quadruplets of four slices (one slice at each position) until a set of 13 quadruplets is formed (13 is the smallest number of repetitions recorded for a given slice position). Out of a given set we extract the probabilities of the three events: 'no feature', 'vortex line' and 'pair of nodal points'. This procedure is repeated 6000 times, where each run provides a new set of 13 quadruplets and the corresponding probabilities $p$, $p_{\rm{VR}}$ and $p_\emptyset$. From the 6000 runs, we obtain the standard deviation of each probability which provides their interval of confidence (error bars in Fig.~\ref{f:stats}).\\

\begin{figure*}[t]
    \begin{center}
    \includegraphics[width=180mm]{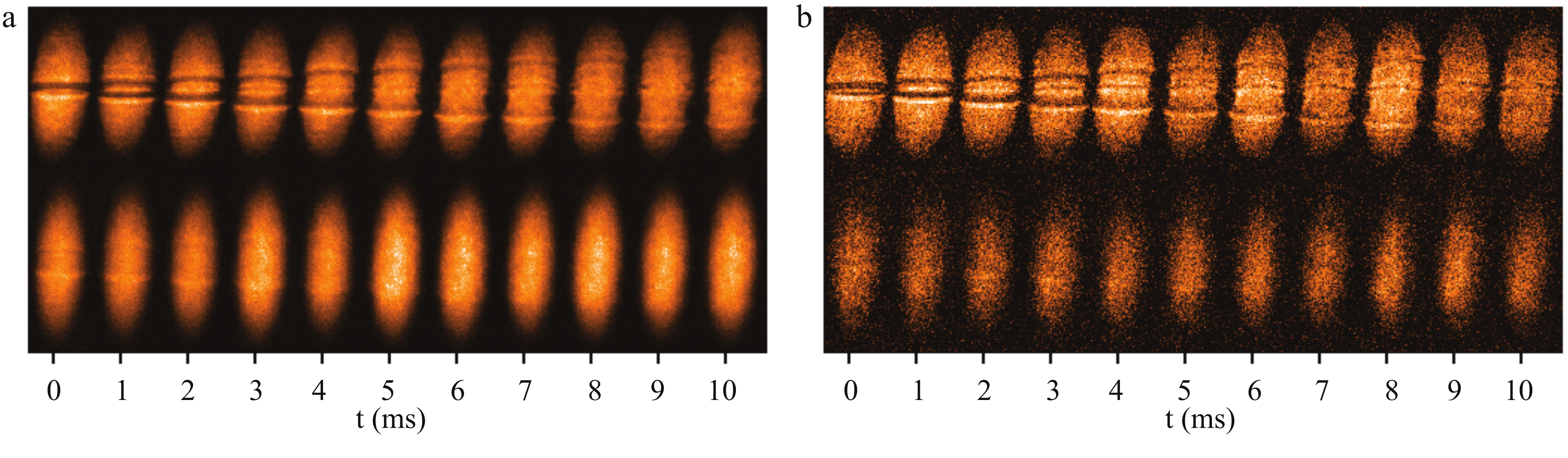}
  \caption{Comparison of images with and without rapid ramp of the magnetic field to the BEC side of the Feshbach resonance. Top: images after 9 ms expansion with the rapid ramp as described previously. Bottom: images after 4 ms expansion without changing the Feshbach field. The expansion time is chosen to approximately match the radial size of the cloud with those obtained with rapid ramp. In each case we show (a) the integrated density (no slicing) and (b) the density of the central layer. Without rapid ramp, the amplitude of the initial wave fronts is strongly reduced.}
    \label{f:SuppFig_RapidRampComparison}
    \end{center}
\end{figure*}

\textbf{Rapid-Ramp versus Time-of-Flight imaging.} The images obtained after the rapid ramp show strong modulations of the density distribution. Based on these images, one might conclude that the phase imprinting leads to an intense disturbance in the density accompanied by shock waves. However, this is not the case. In order to clarify this point, we show, in Fig.~\ref{f:SuppFig_RapidRampComparison}, the time sequence of the dynamics following the imprint by probing the density distribution after a time of flight expansion of 4\,ms, without any change of the scattering length. One observes only weak perturbations in the density profile. In particular, one can see two density wavefronts of small amplitude propagating in opposite direction towards the $z=\pm R_z$ edges of the cloud, that are identified as the fast wavefronts seen in the rapid ramp images. For a more direct comparison, the figure also shows the same time sequence recorded after time of flight combined with the slicing phase, which obviously suffers from a lower signal to noise. This comparison provides evidence that the phase imprint performed in our experiment is indeed a gentle process, which predominantly affects the phase of the superfluid order parameter rather than the density.

\begin{figure*}[h]
    \begin{center}
    \includegraphics[width=175mm]{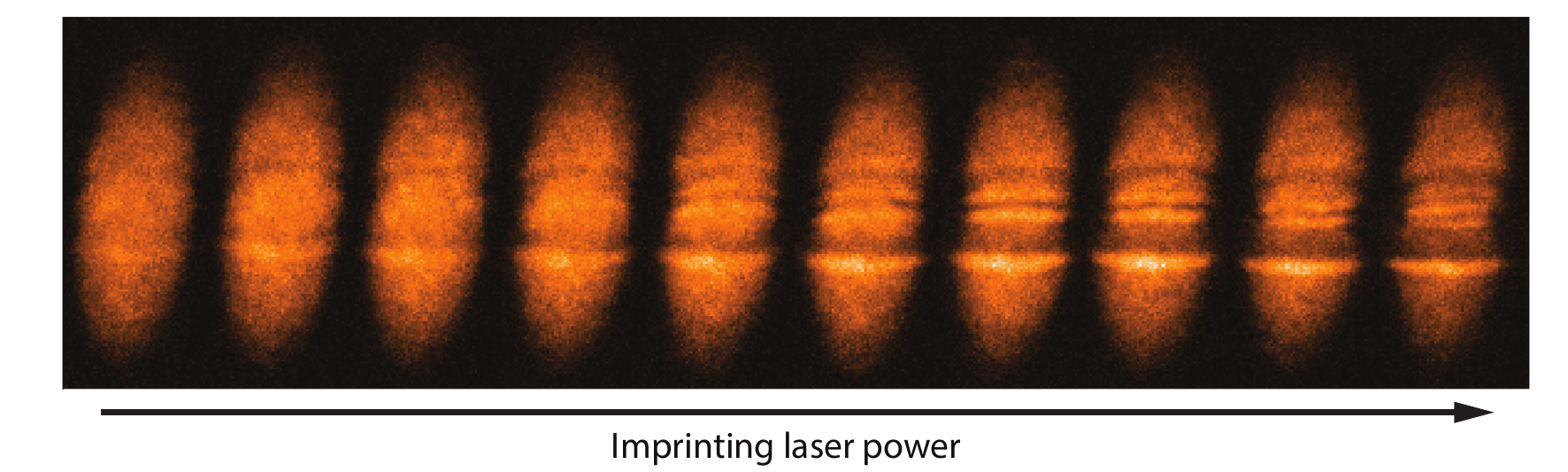}
  \caption{Integrated density after rapid ramp of a phase imprinted superfluid with increasing imprinting power. As the power is increased (images from left to right) we observe the gradual change from a shallow and strongly bent soliton to a straight dark soliton of high contrast to two solitons.}
    \label{f:velocity}
    \end{center}
\end{figure*}

\clearpage

\textbf{Velocity Control of the Soliton.}  By increasing the intensity of the imprinting laser we observe (see Fig.~\ref{f:velocity}) the gradual change from creating fast shallow solitons to slow high contrast solitons and, at even higher power, the formation of multiple solitons. The fast solitons are associated to a strong bending due to the radial inhomogeneity (see the discussion on the dominant snaking mode in the main text). As the imprinting laser intensity directly controls the phase jump across the soliton, this observation constitutes a qualitative demonstration of the soliton's current-phase relation. \\

\textbf{Trapping potential.}
The trapping potential has the following form
%\begin{eqnarray}
%\nonumber V(x,y,z)&=&\frac{1}{2}m\omega^2_z %z^2-\frac{1}{4}m\omega^2_z(x^2+y^2)\\
%\label{eq:fullpotential} &&+V_0\left [
%1-e^{-2\left(\frac{x^2+y^2}{W^2}\right)}
%\right ]+m a y\,.
%\end{eqnarray}
\begin{eqnarray}
\label{eq:fullpotential}  
V(x,y,z)&=&\frac{1}{2}m\omega^2_z z^2-\frac{1}{4}m\omega^2_z(x^2+y^2)\nonumber\\
&+&V_0\left [1-e^{-2\left(\frac{x^2+y^2}{W^2}\right)}\right ]+m a y\,.
\end{eqnarray}

The first term is a saddle point potential from the magnetic field curvature. The second term describes the optical dipole trap (ODT), with a trap depth $V_0/h=11(1)\,\mathrm{kHz}$ and a gaussian waist $W=125(5)\,\mu\mathrm{m}$. The ODT counters the anti-confinement in the radial $x$ and $y$ directions due to the magnetic saddle potential. The effective confinement in the $x$ and $y$ directions, in the absence of gravity, can then be modeled as a harmonic potential with trapping frequency $\omega_{\perp}=\left(\frac{4V_0}{mW^2}-\frac{1}{2}\omega^2_z\right)^{1/2} =2\pi\,69(6)\,\mathrm{Hz}$.
%In our experiment, we have $\omega_z/(2\pi)=10.871(8)\,\mathrm{Hz}$, $V_0=h\times 10.2(2)\,\mathrm{kHz}$, and $W=115(5)\,\mu m$.
Since gravity is not completely compensated, the atoms experience an effective vertical acceleration of $a=1.9(1)\, \mathrm{m/s}^2$. This gradient also shifts the saddle point of the potential in the vertical direction by $10\,\mu\mathrm{m}$.

Expanding the potential for small distances from the origin, as in Ref.~\cite{wlaz14lifecycle}, one obtains
\begin{eqnarray}
\label{eq:approxpotential} V\approx\frac{1}{2}m(\omega^2_zz^2+\omega^2_xx^2+\omega^2_yy^2)+Cy^3+Cx^2y\,,
\end{eqnarray}
with $\omega_x=\omega_{\perp}\sqrt{1-2\delta/3}\simeq\omega_{\perp}(1-\delta/3)$ and $\omega_y=\omega_{\perp}\sqrt{1-2\delta}\simeq\omega_{\perp}(1-\delta)$, where $\delta\equiv\frac{3}{4}\frac{ma^2}{V_0\omega^2_{\perp}}$ and $C\equiv\frac{2m\omega^4_{\perp}}{3a}\delta$ are parameters quantifying the anharmonicity of the potential. $C$ is to be compared to $C_0=\frac{m\omega^2_y}{2R_y}$, where $R_y=61(9)\,\mu\mathrm{m}$ is the Thomas-Fermi radius in the vertical direction. We note that the model of Ref.~\cite{wlaz14lifecycle} does not include the correction to $\omega_x$ and the $Cx^2y$ term. With the given definition for $\delta$ and $C$ and our experimental parameters, one has $\delta=0.019(4)$ and $C=0.16(8)\,C_0$.\\

\begin{figure}
    \begin{center}
    \includegraphics[width=86mm]{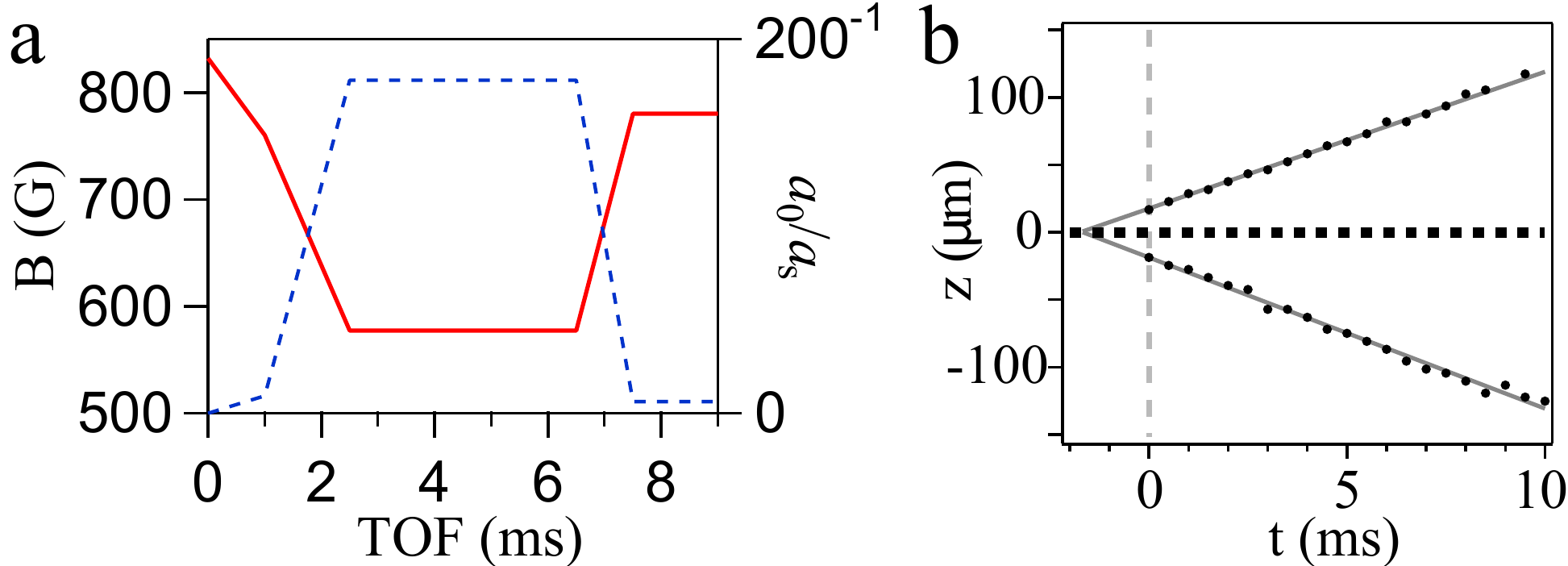}
  \caption{a) Schematics of the rapid ramp. The solid line shows the ramp of the Feshbach field. At wait time $t$ after the imprint, the field is ramped from 832G to 760G in 1 ms, and then ramped down to 576G over 1.5 ms. The field is held at that value for 4 ms, and then ramped back to 780G over 1 ms. The field is held at 780G for 1.5 ms before the spatially selective optical pumping light is applied and the cloud is imaged. The dashed line shows the corresponding $a_0/a_s$, which is 0 at 832G, $4300^{-1}$ at 760G, $225^{-1}$ at 576G, and $6500^{-1}$ at 780G. b) Extraction of the effective time delay after the rapid ramp, from the extrapolation of the sound waves trajectories. Solid black circles: upper and lower sound wave positions as function of time $t$. The linear fit (solid gray lines) yields an effective lag of $\Delta t\approx 1.7$ ms.}
    \label{f:SuppFig_TOFDynamic}
    \end{center}
\end{figure}

\textbf{Dynamics during rapid ramp.} The detection procedure requires a 9\,ms interval between the time $t$ of the dynamics that we want to observe and the time when the image is actually taken. During this phase, the cloud spends $\sim 2\,$ms in the strongly interacting regime $(k_Fa_s)^{-1} \lesssim 1$ where the expansion is hydrodynamic. For the rest of the rapid ramp, the cloud is so weakly interacting that the expansion can be essentially considered ballistic. Therefore, we estimate that the 9\,ms of the combined rapid ramp and expansion can be accounted for as an effective lag of $\sim 2\,$ms in the dynamics of the observed images. One can notice this lag in Fig.~\ref{f:SuppFig_TOFDynamic}b, showing a zoom on the early stage of the sound trajectories. By extrapolating the sound trajectory, we find an effective lag of $1.7$ms, consistent with our estimate.\\

\begin{figure*}
    \begin{center}
    \includegraphics[width=175mm]{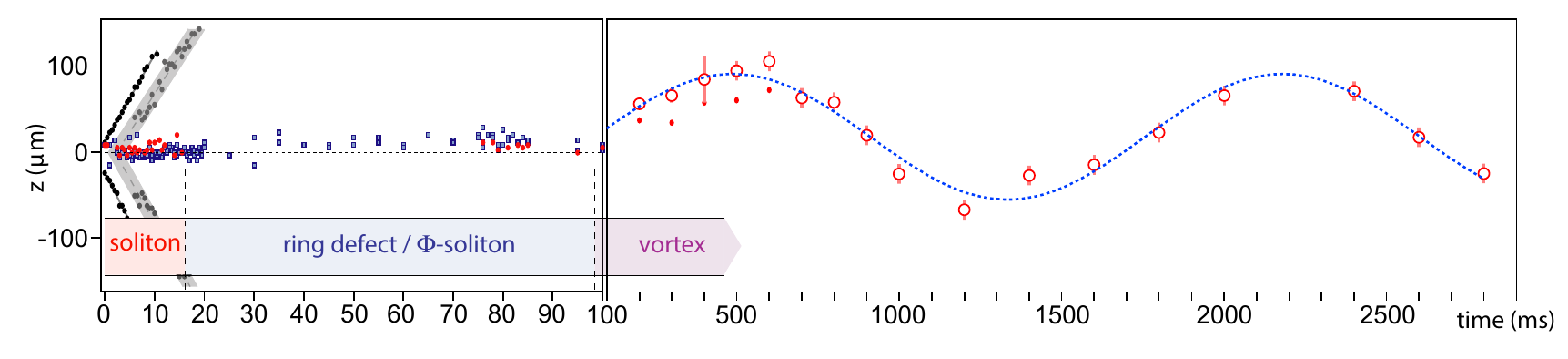}
  \caption{Displacement as a function of time for sound waves and defects. Left panel: location of the sound waves and the defect. This graph is obtained from the residuals displayed in Fig. 4 of the main text. Black solid circles: initial sound waves. Gray solid circles and band: second set of sound waves. Red circles: defect in the central axial cut. Blue squares: defect in the outer axial cuts. Right panel: precession motion (projected on the $z$-axis) of the solitonic vortex resulting from the cascade (red open circles). This graph is obtained from images of line of sight integrated densities. Red solid circles are the locations of the solitonic vortex as detected in the central axial cut of the central slice. }
    \label{f:SuppFig_TrackDefecct}
    \end{center}
\end{figure*}

\textbf{Tracking the waves.}
In Fig.~\ref{f:SuppFig_TrackDefecct}, we show the displacement as a function of time for sound waves and defects, as extracted from the residuals in Fig. 4 of the main text (for $t\leq 100\,\mathrm{ms}$), as well as from the residual of integrated images (for $t\geq\,$100\,ms). This figure provides an overview of the superfluid dynamics following the phase imprint. The various trajectories plotted here show the initial sound propagation, the second set of sound waves, the slowly moving soliton, progressively decaying into a single solitonic vortex, which then precesses in the superfluid. The precession is seen in the right panel as an oscillation along the $z$-axis with a period of $\sim\,1.4\,$s.

\end{document}